\documentclass[10pt,english,aps, notitlepage,onecolumn,superscriptaddress,shortbibliography,nofootinbib,floatfix]{revtex4-2}

\usepackage{mathtools,amsthm,amsmath,amsfonts,amssymb}
\usepackage{lipsum}
\usepackage{subfigure}
\usepackage[colorlinks = true,
            linkcolor = blue,
            urlcolor  = blue,
            citecolor = blue,
            anchorcolor = blue]{hyperref}
\usepackage{physics}
\usepackage{comment}
\usepackage{pgfplots}
    \usetikzlibrary{positioning}
\usepackage[normalem]{ulem}
\usepackage[noend]{algpseudocode}
\usepackage[dvipsnames]{xcolor}
\usepackage{orcidlink}
\usepackage{tikz}
    \usetikzlibrary{math}
\usepackage{parskip}
\usepackage{graphicx, subcaption}
\usepackage{caption}
\usepackage[newcommands]{ragged2e}

\def\be{\begin{equation}}
\def\ee{\end{equation}}

\def\sJ{\mathcal{J}}

\def\bea{\begin{eqnarray}}
\def\eea{\end{eqnarray}}
\def\aa{\hat{\mathfrak{a}}}
\def\qq{\mathfrak{q}}

\def\HH{\hat{H}_0}
\def\nn{\bf{n}}
\def\fid{A}
\def\kk{\hat{k}_0}
\def\Hkryl{\mathcal{H}_{\mathrm{Krylov}}}
\def\td{t_{\text{dep}}}
\def\trev{t_{\text{rev}} }

\DeclareMathOperator{\thetap}{\theta^{\prime}}

\newcommand{\ictpsaifradd}{ICTP South American Institute for Fundamental Research \\
Instituto de F\'{i}sica Te\'{o}rica, UNESP - Univ. Estadual Paulista \\
Rua Dr. Bento Teobaldo Ferraz 271, 01140-070, S\~{a}o Paulo, SP, Brazil
}

\begin{document}

\title{Revival Dynamics from Equilibrium States: Scars from Chords in SYK}

\author{Debarghya Chakraborty\,\orcidlink{0009-0000-1078-5348}}
    \email{debarghya.chakraborty@ictp-saifr.org}
    \affiliation{\ictpsaifradd}

\author{Dario Rosa\,\orcidlink{0000-0001-9747-1033}}
    \email{dario\_rosa@ictp-saifr.org}
    \affiliation{\ictpsaifradd}

\begin{abstract}
We develop a novel framework to build quantum many-body scar states in bipartite systems characterized by perfect correlation between the Hamiltonians governing the two sides. By means of a Krylov construction, we build an interaction term which supports a tower of equally-spaced energy eigenstates. This gives rise to finite-time revivals whenever the system is initialized in a purification of a generic equilibrium state. The dynamics is universally characterized, and is largely independent of the specific details of the Hamiltonians defining the individual partitions. By considering the two-sided chord states of the double-scaled SYK model, we find an approximate realization of this framework. We analytically study the revival dynamics, finding rigid motion for wavepackets localized on the spectrum of a single SYK copy. These findings are tested numerically for systems of finite size, showing excellent agreement with the analytical predictions.
\end{abstract}
\maketitle
\tableofcontents
\section{Introduction}
\label{sec:intro}

The study of how an isolated quantum system relaxes and equilibrates, starting from an out-of-equilibrium configuration, is a question that dates back to the early days of quantum mechanics. Indeed, the unitary nature of time evolution places severe constraints on the ability of a closed quantum system to reach thermal equilibrium. These lines of investigation are becoming even more pressing in recent years, since it is now experimentally possible to build quantum many-body systems in almost perfect isolation from the environment, thus making their dynamics effectively unitary \cite{trotzky2012probing, kaufman2016quantum}.
Nowadays, the thermalization process is understood as a consequence of the celebrated Eigenstate Thermalization Hypothesis (ETH) \cite{deutsch1991quantum, srednicki1994chaos, dalessio2016from}. Making a long story short, thermalization is reached at the level of the expectation values of \textit{local} observables, rather than at the level of quantum states (an idea that was already envisioned by von Neumann). According to the ETH predictions, for a given local observable (\textit{i.e.} an observable acting on a finite number of degrees of freedom), the rest of the system acts as an effective thermal bath, thus making thermalization compatible with unitarity. At the same time, the connection between ETH and Random Matrix Theory (RMT) implies that the eigenstates of a thermalizing Hamiltonian behave as \textit{typical} (or random) states in the many-body Hilbert space, thus following a \textit{volume-law} scaling of the entanglement \cite{dalessio2016from}.

Simultaneously, an active research line explores mechanisms by which quantum many-body systems can escape from ETH and thermalization.
Among these mechanisms, the presence of an extensive number of conserved quantities in involution (the so-called integrable systems) \cite{rigol2007relaxation}, and the presence of disorder (leading to the debated notion of many-body localization \cite{basko2006metal-insulator, abanin2019colloquium, sierant2025many-body}) have been heavily investigated in the past as examples of \textit{strong} ergodicity-breaking, \textit{i.e.} systems that fail to thermalize for \textit{generic} choices of the initial conditions.
On the other hand, quantum many-body scars (QMBS) have emerged as a robust source of \textit{weak} ergodicity breaking in quantum many-body systems \cite{turner2018weak, moudgalya2022quantum, chandran2023quantum}. 
In short, systems with QMBS are characterized by a \textit{generically} thermalizing spectrum.
However, they present a \textit{small} set (typically a set of the same order as the number of degrees of freedom) of \textit{anomalous} (in the sense of showing marked deviations from the RMT predictions) eigenstates which are (at least approximately) equally spaced in energy. 
Although these anomalous eigenstates form a measure-zero subset of the full spectrum, they can deeply affect the dynamics of \textit{specific} initial states. The reason is easy to guess: if the system is initially prepared in a state having \textit{large} (or even complete) overlap with the special eigenstates, its dynamics will be highly non-ergodic and will show huge revivals, with the period proportional to the inverse of the energy gap among the equally spaced eigenstates. 

Obviously, this ergodicity-breaking mechanism is of any practical interest only if the initial states showing huge overlaps with the anomalous eigenstates are reasonable enough to be prepared and characterized. This latter condition is usually ensured by the property of the QMBS of being low-entangled.

In this paper, we will develop a novel mechanism to build a tower of special eigenstates which are equally spaced in energy, \textit{i.e.} a new mechanism of building QMBS and non-ergodic dynamics consequently. The construction is quite general and flexible, taking as a starting point a bipartite system, \textit{i.e.} a system whose degrees of freedom can be separated into two groups (conventionally called ``left'' and ``right''), with the uncoupled dynamics governing the two parties being perfectly \textit{anticorrelated} so that, loosely speaking, the right system is controlled by the same Hamiltonian of the left system but with all the couplings having the sign flipped (in the case that $\hat{H}_L$ enjoys particle-hole symmetry, the request of perfect anticorrelation can be relaxed and replaced with a request of perfect \textit{correlation}). This notion will be made more precise later, but schematically we can write $\hat{H} = \hat{H}_L - \hat{H}_R$, with $\hat{H}_R \sim \hat{H}_L$.  

Given this setup, it is immediately clear that \textit{every} diagonal eigenstate, \textit{i.e.} every state having the schematic form $\ket{\psi} \equiv \sum_{n} c_n \ket{E_n}_L \otimes \ket{E_n}_R$, with $\ket{E_n}_{L}$ ($\ket{E_n}_{R}$) being the eigenstates of the left (right) Hamiltonian, is a zero-energy eigenstate for the full Hamiltonian and, given its special left-right correlations, it can be called a scarred eigenstate. Among those states, the infinite temperature thermofield double state (TFD), denoted by $\ket{0}$ and called ``rainbow state'' in \cite{langlett_2022}, will play a very special role. 
Up to now, the construction has been quite boring: we have taken a bipartite fully anticorrelated system and observed that it admits a large set of zero modes. To make things more interesting, we aim to introduce an explicit coupling between the two halves, in such a way that $\ket{0}$ will be part of a tower of equally spaced in energy special eigenstates, thus supporting non-trivial non-ergodic dynamics. Towards this goal, following a standard Krylov-like approach, we define a subspace of the total Hilbert space, called $\mathcal{H}_\mathrm{Krylov}$, which is obtained by considering all the integer powers of the left Hamiltonian, $\hat{H}_L$, acting on $\ket{0}$, \textit{i.e.} we define $\mathcal{H}_\mathrm{Krylov} \equiv \mathrm{span}\left\{\hat{H}_L^k \ket{0}, \forall k \in \mathbb{N}\right\}$. Obviously, $\mathcal{H}_\mathrm{Krylov}$ forms a zero-energy subspace of the total Hamiltonian and its states, after a partial trace removing the right (left) copy, are \textit{equilibrium} states (and therefore non-dynamical) for the uncoupled left (right) system, \textit{i.e.} they are diagonal once written in the $\left\{\ket{E_n}_L \right\}$ ($\left\{\ket{E_n}_R \right\}$) basis. The last task is to introduce a non-trivial (but non-ergodic) dynamics in $\mathcal{H}_\mathrm{Krylov}$ via a suitable \textit{grading operator}, $\hat{n}_0$, which acts on $\mathcal{H}_\mathrm{Krylov}$ as an endomorphism splitting it into subspaces characterized by different (integer) eigenvalues. To this end, we notice that, via a standard Gram-Schmidt procedure, the very structure of $\mathcal{H}_\mathrm{Krylov}$ provides a natural grading operator: after orthonormalization, $\mathcal{H}_\mathrm{Krylov}$ can be naturally written as $\mathcal{H}_\mathrm{Krylov} = \mathrm{span}\left\{\ket{0}, \ket{1}, \dots , \ket{\mathrm{dim} (\mathcal{H}_\mathrm{Krylov})}\right\}$, from which we obtain $\hat{n}_0 = \sum_{n} n \ket{n} \bra{n}$.  

In summary, the deformed Hamiltonian $\hat{H}(\mu) = \hat{H} + \mu \hat{n}$, with $\mu$ being a coupling constant and $\hat{n}$ being a generic embedding of the operator $\hat{n}_0$ into the full Hilbert space, by construction induces non-ergodic dynamics, characterized by finite-time revivals proportional to $1/\mu$, for \textit{every state} living in $\mathcal{H}_\mathrm{Krylov}$. The features of this induced non-ergodic dynamics, described in Sec.~\ref{sec:general_construction} are particularly interesting once described from the point of view of an observer living on the left (or, equivalently, the right) system only. As we already remarked, the states living in $\mathcal{H}_\mathrm{Krylov}$, once viewed from the left system (equivalently, after the partial trace removing the right system) and in the absence of the extra interaction, are equilibrium states. Therefore, once the coupling between the two systems is turned on, from the point of view of the left system the dynamics is turning into a dissipative dynamics (with the role of the ``environment'' played by the right copy) but highly non-ergodic, in such a way the system keeps evolving from equilibrium states into equilibrium states and periodically returning to the original configuration. Using an analogy, the dynamics experienced by one of the two copies shares some similarities with a hypothetical block of ice that, left free in the environment, first increases its temperature and melts down into water, but later reduces its temperature again and rebuilds itself. 

A major obstacle in realizing the construction outlined so far relies on finding specific examples in which the operator $\hat{n}_0$, and the subsequent embedding $\hat{n}$, takes a natural form in terms of the microscopic operators defining the left and the right systems. This is a non-trivial problem, since $\hat{n}_0$ is only implicitly defined via the moments of the left Hamiltonian, \textit{i.e.} via quantities like $\tr \left(\hat{H}_L^{k}\right)$. While we leave the investigation of the general conditions ensuring a natural form for the operator $\hat{n}_0$ for the future, in this paper we will show an important example where an \textit{approximate} realization of the operator $\hat{n}$ can be explicitly described. We will see in Sec. \ref{sec:approx_realization_SYK} that this approximate setup can be realized when the left and right systems are described by the Sachdev-Ye-Kitaev (SYK) model \cite{Maldacena_16, polchinski2016the}, an all-to-all system of Majorana fermions with the interaction terms given by quenched disorder variables. 

According to the rules of the construction, we will assume that both the left and the right Hamiltonians are described by \textit{exactly} the same disorder realization (up to a possible global minus sign, as we will describe). We will then show that, in the so-called \textit{double-scaling limit} of the theory -- a limit in which both the number of fermions, $N$, and the degree of the interaction, $p$, are sent to infinity in such a way that the ratio $\lambda = \frac{2 p^2}{N}$ is kept fixed -- the space $\mathcal{H}_\mathrm{Krylov}$ is identified with a sector of the two-sided \textit{chord Hilbert space} of the double-scaled SYK model \cite{Berkooz_2018, Berkooz_2019, Lin_2022, Berkooz:2024lgq}. In this context, $\hat{n}_0$ reduces to the \textit{chord number} operator, for which an approximate realization can be given in terms of the \textit{size operator} $\hat{S}$, a bilinear of the left and right Majorana operators \cite{Lin_2022}. The total Hamiltonian is related to the Maldacena-Qi traversable wormhole \cite{maldacena2018eternaltraversablewormhole} up to the relative sign. The two models show drastic physical differences \cite{Chakraborty_2026}, a fact which can be attributed to the holographic interpretation of the two variants of the Hamiltonian, in terms of generators of $AdS_2$ isometries. The choice that $\hat{H}(\mu)$ at $\mu=0$ annihilates the thermofield double, can be interpreted as the boost isometry of the eternal black hole \cite{Maldacena_2003}.

Given the high degree of analytical control at our disposal for the double-scaled SYK model, in Sec.~\ref{sec:dynamics_double_scaling} we will investigate in detail the dynamics obeyed by the states living in $\mathcal{H}_\mathrm{Krylov}$. A notable result is that any thermofield double state evolves into states that, to a very good approximation, remain thermal when reduced to the left (or right) copy, with a time-dependent inverse temperature $\beta_{\text{eff}}(t)$ that can be computed analytically. Moreover, we will show that a tensor product of eigenstates of left and right systems, \textit{i.e.} a state of the form $\ket{E}_L \otimes\ket{E}_R$, evolves like a \textit{sharply localized particle} along the spectrum of the decoupled system. This result gives an extreme manifestation of the highly non-ergodic dynamics enjoyed by $\mathcal{H}_{\mathrm{Krylov}}$. Finally, we will investigate the persistence of these dynamical features outside of the double-scaling limit, both analytically, by considering the $\lambda \to 0$ limit in Sec.~\ref{sec:continuum_limit}, and numerically at finite values of $N$ and $p$ in Sec.~\ref{sec:finite_size_revivals}. In particular, the numerical results obtained at finite sizes will show excellent agreement with the analytical predictions, thus providing strong evidence for the robustness of the entire construction.

\subsection*{Connections with related constructions}
The idea of building quantum many-body scars in bipartite setups is not new, and other approaches have already been considered in the literature, in particular in \cite{langlett_2022} and \cite{Dong_2023}. Nevertheless, these approaches differ significantly from the construction presented in this paper. We will now make some comments to relate the results of \cite{langlett_2022} and \cite{Dong_2023} to the approach of this paper. 

In \cite{langlett_2022}, the authors focus their attention on the rainbow state $\ket{0}$, \textit{i.e.} on the very first state of the chord Hilbert space, and they consider a coupling term, $\hat{V}_c$, for which $\ket{0}$ is an eigenstate as well. Starting from this setup, they consider situations where they can find an operator $\hat{\mathcal{O}}$, such that it commutes with $\hat{H}_L - \hat{H}_R$ and it satisfies with $\hat{V}_c$ a relation of the form $\left[\hat{V}_c, \, \hat{\mathcal{O}} \right] \propto \hat{\mathcal{O}}$.
In this way, they can build other quantum many-body scar states on top of the rainbow state, by considering the states of the form $\hat{\mathcal{O}}^k \ket{0}$. The construction of \cite{Dong_2023} is somewhat similar in spirit, but including a tower of states constructed from $\hat{H}_L \ket{0}$, \textit{i.e.} the state $\ket{1}$ in the Krylov space in addition to $\ket{0}$. All these approaches create \textit{exact} many-body scars while imposing extra conditions on the left and right Hamiltonians. We can say that the approach presented in this paper is \textit{orthogonal} to them: instead of adding extra structures to the left and right Hamiltonians, we consider the full tower of states obtained from $\ket{0}$ by the Krylov construction. 

\section{The general construction} 
\label{sec:general_construction}
Consider a bipartite system, in which both the partitions -- denoted as ``left'' (L) and ``right'' (R), respectively -- are made of $N$ sites, with each site consisting of a local Hilbert space of dimension $d$, so that the total Hilbert space $\mathcal{H} = \mathcal{H}_L \otimes \mathcal{H}_R$ has dimension $d^{2N}$. 
Let us also define a bijective map -- called \textit{mirror map} in \cite{langlett_2022} and denoted by $\mathcal M$ -- realizing a one-to-one correspondence between the sites on the left copy, denoted by $i$, and the sites on the right copy, denoted by $\Tilde{i}$, \textit{i.e.} $\mathcal{M}: i \to \tilde i, \ \forall i = 1, \dots, N$.

Given a certain Hamiltonian, $\hat{H}_L$, acting on the left system only, we define a corresponding Hamiltonian acting on the right system, $\hat{H}_R$, as follows:
\begin{equation}
    \label{eq:H_R_def_gen}
    \hat{H}_R \equiv \mathcal{M}\hat{H}_L^\ast \mathcal{M} ,
\end{equation}
and we construct the total Hamiltonian as \footnote{In the following, with a slight abuse of notation, we will denote with $\hat{H}_L$ ($\hat{H}_R$) \textit{both} the operator acting on $\mathcal{H}_L$ ($\mathcal{H}_R$) and the operator acting on the full Hilbert space $\mathcal{H}$, obtained from  $\hat{H}_L$ ($\hat{H}_R$) by taking the tensor product with the identity operator acting on the right (left) copy.}
\begin{equation}
    \label{eq:total_hamiltonian_no_interaction_def}
    \hat{H} = \hat{H}_L \otimes \hat{\mathbb{I}} - \hat{\mathbb{I}} \otimes \hat{H}_R .
\end{equation}

It can be immediately realized that $\hat{H}$ has the following zero-energy eigenstate
\begin{equation}
    \label{eq:rainbow_state_def}
    \ket{0} \equiv \frac{1}{d^{N/2}} \sum_{n = 1}^{d^{N}} \ket{E_n} \otimes \ket{\mathcal{M}E_n^\ast} ,
\end{equation}
where $\ket{E_n}$ are the eigenstates of the left Hamiltonian, \textit{i.e.} $\hat{H}_L \ket{E_n} = E_n \ket{E_n}$, and $\ket{\mathcal{M}E_n^\ast} \equiv \left( \mathcal{M} \ket{E_n}\right)^\ast$. From Eq.~\eqref{eq:rainbow_state_def} it follows immediately that $\ket{0}$ constitutes an infinite-temperature thermofield-double state (TFD) for $\hat{H}_L$. The entanglement properties of $\ket{0}$ have been studied at length in \cite{Maldacena_2013,Ram_rez_2015,langlett_2022, Schuster_2022} and, due to the peculiar shape of its left-right correlations, it has been called the \textit{rainbow state}. 

Let us now consider the following subspace of $\mathcal{H}$:
\begin{equation}
    \label{eq:H_TFD_def}
    \mathcal{H}_{\mathrm{Krylov}} \equiv \mathrm{span} \left\{\hat{H}_L^k\ket{0}, \ \forall k \in \mathbb{N}\right\}  .
\end{equation}
Given the property $\left(\hat{H}_L - \hat{H}_R \right)\hat{H}_L^k\ket{0} = 0$, valid for any $k$, it follows that \textit{every state} in $\mathcal{H}_\mathrm{Krylov}$ is a zero-energy eigenstate of $\hat{H}$. Moreover, consider a generic finite-temperature TFD state 
\begin{equation}
    \label{eq:TFD_finite_T_def}
    \ket{\beta} \equiv \frac{1}{\sqrt{Z(\beta)}}  \sum_{n = 1}^{d^N} e^{-\frac{\beta}{2} E_n} \,\ket{E_n} \otimes \ket{\mathcal{M}E_n^\ast} ,
\end{equation}
with $\beta \equiv 1/T$ being the inverse temperature (we work in units with $\hbar = k_B = 1$) and $Z(\beta) = \sum_{n} e^{- \beta E_n}$ the partition function at inverse temperature $\beta$. It is immediate to verify that $\ket{\beta} \in \mathcal{H}_\mathrm{Krylov} \ \forall \beta$, \textit{i.e.} every finite temperature TFD belongs to $\mathcal{H}_\mathrm{Krylov}$ and therefore it is a zero-energy eigenstate of $\hat{H}$. Notice that $\mathcal{H}_\mathrm{Krylov}$ includes a much wider class of states than just the TFDs. Assuming that $\hat{H}_L$ does not have degenerate eigenvalues, all states of the form $\sum_n c_n \ket{E_n} \ket{\mathcal{M}E_n^\ast}$ are contained in $\mathcal{H}_\mathrm{Krylov}$. By construction, $\hat{H}_L$ and $\hat{H}_R$ have identical action on $\mathcal{H}_\mathrm{Krylov}$ and keep it invariant. Therefore, on $\mathcal{H}_\mathrm{Krylov}$, we define  $\HH \equiv \hat{H}_L = \hat{H}_R$ 

Given this premise, let us assume that we can find an operator $\hat{n}$ on $\mathcal H$, such that when acting on the elements of $\mathcal{H}_\mathrm{Krylov}$ it satisfies:
\begin{enumerate}
    \item $\hat{n}$ restricted to $\mathcal{H}_{\mathrm{Krylov}}$ is an endomorphism  \textit{i.e.} $\hat{n} \ket{\psi} \in \mathcal{H}_{\mathrm{Krylov}} \ \forall \ket{\psi} \in \mathcal{H}_{\mathrm{Krylov}}$. We denote by $\hat{n}_{0}$ the restriction of $\hat{n}$ on the elements of $\mathcal{H}_{\mathrm{Krylov}}$, \textit{i.e.} $\hat{n}_0 \equiv \left. \hat{n} \right|_{\mathcal{H}_{\mathrm{Krylov}}}$.
    \item $\hat{n}_{0}$ is a \textit{counting operator}, \textit{i.e.} its eigenvalues can be labeled by integers and they are equally spaced. We denote the eigenstates of $\hat{n}_0$ as $\ket{n}$.
\end{enumerate}
Under these conditions, it becomes natural to modify the Hamiltonian to be
\begin{equation}
    \label{eq:total_hamiltonian_mu-shifted}
    \hat{H} \to \hat{H}(\mu) = \hat{H}_L \otimes \hat{\mathbb{I}} - \hat{\mathbb{I}} \otimes \hat{H}_R + \mu \hat{n} ,
\end{equation}
since this ensures that \textit{all} the eigenstates of $\hat{n}_0$, $\ket{n}$, form a special set of eigenstates of $\hat{H}(\mu)$ living in $\mathcal{H}_{\mathrm{Krylov}}$, as they satisfy the property 
\begin{equation}
    \label{eq:n_eigenstates_as_many-body_scars}
    \hat{H}(\mu) \ket{n} \propto (\mu n) \ket{n} ,
\end{equation} 
\textit{i.e.} they are \textit{many-body scars} for $\hat{H}(\mu)$.

However, without further constraints, it is not possible to exclude cases in which $\mathcal{H}_\mathrm{Krylov}$ still forms a non-dynamical subsector of $\mathcal{H}$. As an example, consider the case in which $\hat{n}_0 \propto \hat{\mathbb{I}}_0$. 
To rule out such a possibility, we want to consider cases where $\hat{n}$ divides $\mathcal{H}_\mathrm{Krylov}$ into distinct eigenspaces whose number scales as a function of $N$. A constructive way of achieving this is to consider $\hat{n}_0$ to be the Krylov number operator, $\hat{n}_\mathrm{Krylov}$, obtained from the Lanczos procedure underlying the explicit construction of an orthonormal basis for $\mathcal{H}_{\mathrm{Krylov}}$. 
In formulas, we consider the following procedure
\begin{align}
    \label{eq:n_0_krylov_explicit}
    \mathcal{H}_\mathrm{Krylov} &\equiv \mathrm{span}\left\{\hat{H}_L^k \ket{0}, \ \forall k \in \mathbb{N} \right\} \Rightarrow \mathrm{span}\left\{\ket{n}, \ \forall n = 1, \dots, \mathrm{dim}(\mathcal{H}_\mathrm{Krylov}), \ \mathrm{s.t.} \ \langle n | m \rangle = 0 \ \forall n\neq m \right\}, \nonumber \\
    &\Rightarrow \hat{n}_0 \equiv \sum n \ket{n}\bra{n},
\end{align}
which guarantees $\text{dim} (\mathcal{H}_\mathrm{Krylov}) $ distinct eigenspaces. Moreover $\hat{H}_0$ becomes tridiagonal in the $\hat{n}_0$ eigenbasis.  
Using this construction, we see that \textit{any} state in $\mathcal{H}_{\mathrm{Krylov}}$ will show non-ergodic dynamics characterized by revivals. In particular, this will include non-ergodic dynamics for \textit{any} TFD state $\ket{\beta}$, thereby realizing an example of weakly broken ergodicity.

The considerations above can be formalized by including this extra requirement on $\hat{n}_0$:
\begin{enumerate}
    \item[$3$.] $\hat{H}_0$, $\hat{n}_0$, and $\ket{0}$ satisfy the following relations
    \begin{align}
        \label{eq:H_0_tridiagonal}
        & \hat{H}_0  = \hat{L}_{+} +\hat{L}_0 + \hat{L}_-, \qquad \mathrm{with} \ \hat{L}_{\pm} \ket{n}  \propto \ket{n \pm 1},  \ \hat{L}_0 \ket{n} \propto \ket{n}   \nonumber  \\
        & \hat{n}_{0}\ket{0} = 0 \, .
    \end{align}
\end{enumerate}
When $\hat{H}_L$ has a symmetric spectrum i.e $\rho(E)$ is an even function about some $E_{*}$, then $\hat{L}_0$ can be made to vanish identically via a shift. Even when this property is not satisfied, it does not affect the scar dynamics, and therefore, we will set it $0$ from now on. 

We note that this construction is similar in spirit to a restricted spectrum generating algebra \cite{Moudgalya_2020, Iadecola_2020, Mark_2020} with the ladder operators being
\be  
\hat{L}_{\pm}  = \frac{1}{2} \big( \hat{H}_0 \mp  2 i \kk   \big), ~~~~~~ \kk \equiv \frac{i}{2} [ \hat{n}_0, \hat{H}_0] ~.
\label{eq:H_and_k_in_chord_basis}
\ee  
The factor of $2$ in the definition of $\kk$ is included to relate it more naturally with another quantity that we will define in Sec \ref{sec:continuum_limit}.  
It is also worth noticing that the above construction does not necessarily require taking $\ket{0}$ as the reference state for the definition of $\mathcal{H}_\mathrm{Krylov}$, and that any TFD state $\ket{\beta}$ can be used instead. The operator $\hat{n}_{0}$ gets modified accordingly. This generalization can be pertinent when studying systems that do not have a good limit for the canonical ensemble at $\beta=0$.  
\subsection{Non-ergodic dynamics: general considerations}
\label{sec:general_non_ergodic_dynamics}
Given $\hat{H}_L$, the reference state $\ket{0}$ and imposing the condition $3.$, the action of the Hamiltonian $\hat{H}(\mu)$ within the subspace $\mathcal{H}_{\mathrm{Krylov}}$ gets completely fixed. Therefore, many aspects of the dynamics in $\mathcal{H}_{\mathrm{Krylov}}$ can be deduced without an explicit knowledge of how $\hat{n}_0$ is extended to the operator $\hat{n}$ acting on the entire space. \footnote{
But certainly, whether $\hat{H}(\mu)$ makes sense physically or not depends on this extension.} In this subsection, we will describe some \textit{universal} features of the dynamics in the scar subspace, which can be deduced without specifying $\hat{H}_0$ in terms of $\hat{L}_{\pm}$ and the closure of their algebra. 

To this end, let us assume that $\hat{H}_{L}$ is a generic non-integrable Hamiltonian satisfying ETH. For the sake of simplicity, we will also assume that $\hat{H}_{L}$ does not have any unitary symmetry, although it is not difficult to extend this discussion in the presence of a few symmetry sectors. For any state $\ket{\psi(0)} = \sum_n c_n \ket{n} \in \mathcal{H}_{\mathrm{Krylov}}$, we have 
\be 
\label{eq:survival_amplitude}
\ket{\psi(t)} \equiv e^{-i \hat{H}(\mu) t } \ket{\psi(0)} = e^{-i \mu \hat{n} t} \ket{\psi(0)}, ~~~ \fid(t)= \bra{\psi(0)} \ket{\psi(t)} = \sum_n \abs{c_n}^2 e^{- i \mu n t}
\ee 
which therefore implies a revival dynamics for the return probability $F(t) \equiv \abs{ \fid(t)}^2$. We call the period of the revivals $t_{\text{rev}}$, and clearly, $t_{\text{rev}} = 2 \pi/{\mu}$.

Let us now analyze the consequences of the revival dynamics for the time evolution of an observable $\hat{O}$ in the Heisenberg picture. We have
\be
\bra{\psi(0)} e^{i \hat{H}(\mu) t } \hat{O} e^{-i \hat{H}(\mu) t } \ket{\psi(0)}  = \bra{\psi(0)} e^{i \mu \hat{n} t}  \hat{O} e^{-i \mu \hat{n} t}  \ket{\psi(0)} ~. \label{eq:Heisenberg_EOM} 
\ee
Given Eq.~\eqref{eq:Heisenberg_EOM}, we readily obtain the universal oscillatory behavior of the expectation values of $\HH$ and $\kk$, valid without the need to specify the details of $\ket{\psi(0)} \in \mathcal{H}_{\mathrm{Krylov}} $  
\bea 
\langle \HH (t) \rangle  &= \langle \HH \rangle \cos \mu t  + 2 \langle  \kk \rangle  \sin \mu t  \label{eq:Heisenberg_H} \\  
2 \langle \kk (t) \rangle  &= 2 \langle \kk \rangle \cos \mu t  - \langle \HH \rangle  \sin \mu t \label{eq:Heisenberg_k} \eea
where $\langle , \rangle$ is a shorthand for the expectation value evaluated in $\ket{\psi(0)}$. To obtain the oscillatory behavior above, we made use of the relation $[\hat{n}_0, \hat{L}_{\pm}] = \pm \hat{L}_{\pm} $ which implies $e^{i \mu \hat{n} t}  \hat{L}_{\pm} e^{-i \mu \hat{n} t} = \hat{L}_{\pm} e^{\pm i \mu t}$. 

Taken together, Eq.~\eqref{eq:Heisenberg_H} and Eq.~\eqref{eq:Heisenberg_k} reproduce the time-evolution generated by a harmonic oscillator Hamiltonian in the scar subspace. $\HH(t)$ plays the role of the position operator, and $\kk(t)$ plays the role of its time derivative, \textit{i.e.} it can be interpreted as the momentum operator. This is precisely because of the way $\HH$ and $\kk$ are expressed in terms of the ladder operators in Eq.\eqref{eq:H_0_tridiagonal} and Eq.~\eqref{eq:H_and_k_in_chord_basis}. Notice that the commutation relation between $\hat{L}_{+}$ and $\hat{L}_{-}$ does not need to be specified. In the concrete example of the SYK model that we study in Section \ref{sec:SYK}, $\hat{L}_{\pm}$ and $\hat{n}_0$ satisfy the Arik-Coon $\qq$-deformed algebra. We also observe that $\langle \HH (t) \rangle^2 + 4 \langle \kk (t) \rangle^2$ and  $\langle \HH (t)^2 + 4  \kk (t)^2 \rangle $ are conserved on the subspace, the latter because $\hat{L}_{+} \hat{L}_{-} + \hat{L}_{+}\hat{L}_{-} =  f(\hat{n}_0)$ is diagonal.  

We stress that $\langle\HH(t) \rangle$ \textit{should not be confused} with the total energy of the coupled system, which, on the contrary, is given by $\langle\hat{n}_0\rangle$ since the contributions of $H_L$ and $H_R$ cancel by construction. To understand the physical meaning of Eq.~\eqref{eq:Heisenberg_H} it is convenient to expand the state $\ket{\psi (t)}$ in the eigenbasis of $\hat{H}_L$ and $\hat{H}_R$ as
\be  
\ket{\psi(t)} = \sum_{k} c_{k}(t) \ket{E_k} \ket{\mathcal{M}E_k^\ast} ~. 
\label{eq:Psi_decomposition_LR_energy_basis}
\ee 
Rewriting Eq.~\eqref{eq:Heisenberg_H} as $\langle \HH(t) \rangle = \sqrt{\langle \HH \rangle^2 + 4\langle \kk \rangle^2 } \cos(\mu t - \arctan( \frac{2 \langle\kk\rangle}{\langle\HH\rangle} ) )$, we see that the mean ``position'' $\langle \HH(t) \rangle = \sum_k \abs{c_k(t)}^2 E_k $ oscillates, taking values $ n \sqrt{\langle \HH \rangle^2 + 4\langle \kk \rangle^2 }$ at $t = \frac{\pi + 2 \arctan( \frac{2 \kk}{\HH} ) }{2 \mu} -  \frac{n \pi}{2 \mu} $, for $n \in \{-1, 0, 1\}$. 

More information about the dynamics, as seen by the left/right system, can be obtained by considering the energy variance. The energy variance $\langle \HH^2 (t)\rangle -\langle \HH(t) \rangle^2$ at arbitrary times can be bounded in terms of the variance of $\hat{k}$ and $\HH$ at $t=0$. For initial states with small variances in $\HH$ and $\kk$ at $t=0$, we see that the non-ergodic dynamics induced by $\mu \hat{n}$ corresponds to a \textit{quasi-rigid} motion of a \textit{particle on the spectrum}, analogous to the dynamics of a coherent state in a harmonic oscillator. More generally, the time-dependent $c_{k}(t)$ in Eq.~\eqref{eq:Psi_decomposition_LR_energy_basis} describes a wavepacket that is initially concentrated, say, to the ``left'' ($\langle \HH \rangle< 0$) half of the spectrum of the decoupled Hamiltonian $\hat{H}_L$ (or $\hat{H}_R$). The wavepacket moves to the ``right'', such that one of the system heats up, until reaching a maximum energy and then bouncing back along the spectrum.  

We will illustrate these ideas precisely by considering different classes of initial states in Section \ref{sec:SYK}. A remarkable consequence of the $\qq$-deformed algebra in DSSYK, is that a perfectly localized particle on the spectrum does not disperse; which means that if $c_{k}(t)$ is a $\delta$ function localized at $x(0)$ at $t=0$, under time evolution it will become a $\delta$ function at  $x(t)$.  
Tracing out one of the copies in Eq.~\eqref{eq:Psi_decomposition_LR_energy_basis} results in the density operators: 
\be  
\rho_L(t) =  \sum_j \abs{c_{j}(t)}^2  \ket{E_j}\bra{E_j} ~~~, \rho_R(t) =\sum_j \abs{c_{j}(t)}^2  \ket{\mathcal{M}E_j^\ast}\bra{\mathcal{M}E_j^\ast} ~.
\label{eq:one-sided} 
\ee 
In the absence of the $\mu \hat{n}$ term in the Hamiltonian, which couples the left and right systems, these would obviously be equilibrium states for the two systems in isolation.
Now, from the point of the left system, the left/right dynamics can be interpreted as a kind of \textit{dissipative dynamics}, with the right system playing the role of the environment. Therefore, it follows that, from the point of view of the left (right) system, we begin in an equilibrium state, which, thanks to the dissipation, develops a harmonic oscillator-like dynamics among equilibrium states, with finite-time revivals in its energy content. The dynamics described via this construction shares a certain similarity with the boundary time-crystal introduced in \cite{Iemini_2018}.

Eq.~\eqref{eq:one-sided} shows that the entanglement entropy between the left and right systems $S(t) \equiv S(\rho_L(t))$ is precisely the Shannon entropy of the probability density along the left (right) energy eigenbasis. Thus $S(t)$ is another measure of the spread of the wavepacket in Eq.~\eqref{eq:Psi_decomposition_LR_energy_basis}. Given that the density operators in Eq.~\eqref{eq:one-sided} are insensitive to the phases of $c_{j}(t)$, it may appear that these relative phases do not play any role in the dynamics. However, operators coupling the left and the right system are sensitive to the phases. The momentum $\kk$ is an example of such an operator, see Sec.~\ref{sec:SYK} for an explicit realization.

Eq.~\eqref{eq:Heisenberg_H} and Eq.~\eqref{eq:one-sided} can be further used to constrain the time evolution of a generic local observable $\hat{O}_L$, defined on the left system and obeying ETH. To start with, we notice that, given the peculiar dynamics obeyed by states belonging to $\mathcal{H}_{\mathrm{Krylov}}$, the time-evolved expectation value $\langle \hat{O}_L (t) \rangle$ is fully determined by the diagonal matrix elements $\bra{E_k}\hat{O}_L\ket{E_k}$. Assuming the validity of ETH, these are expressed in terms of the smooth function $O_L (E)$. 
For configurations which keep a small variance in $\HH$ under time evolution, we can further assign a time-dependent effective temperature $\beta_{\text{eff}}(t)$ determined by 
\be 
\frac{\tr(\hat{H}_L e^{- \beta_{\text{eff}}(t) \hat{H}_L } )}{\tr(e^{- \beta_{\text{eff}}(t) \hat{H}_L } ) }  = \langle \HH(t) \rangle, 
\label{eq:beta_effective}
\ee
where the right hand side is determined by Eq.~\eqref{eq:Heisenberg_H}, and therefore, using the properties of ETH we get,
\be 
\langle \hat{O}_L (t) \rangle =  \frac{\tr(\hat{O}_L e^{- \beta_{\text{eff}}(t) \hat{H}_L } ) }{\tr(e^{- \beta_{\text{eff}}(t) \hat{H}_L } ) }.
\label{eq:O_effective}
\ee 
Thus, the knowledge of \textit{equilibrium} expectation values for $\hat{O}_L$ in the left system, determines its  dynamics under the coupled Hamiltonian. There is a strong formulation of the ETH which says that the reduced density matrix of eigenstates restricted to small subsystems is thermal to a high precision; see \cite{Garrison_2018, Dymarsky_2018} for the precise statement. In our setting, this means that the temperature dependence of Eqs.~\eqref{eq:beta_effective} and \eqref{eq:O_effective} extends to the reduced density matrix of small subsystems. In particular, the time-dependent entanglement entropy density of small subsystems is readily identified with the thermal entropy density $ s_{\text{th}}(\beta_{\text{eff}}(t))$.

If $Z(\beta)$ is known, then it is possible to solve for $\beta_{\text{eff}}(t)$, using Eqs.~\eqref{eq:beta_effective} and~\eqref{eq:Heisenberg_H}. An important example, relevant to the explicit construction presented in Section~\ref{sec:revival_dynamics_DSSYK}, is when the spectral density for the left system, $\rho(E)$, is approximately Gaussian. This is widely relevant for the high temperature limit of spin systems with few-body interactions. In this case the LHS of Eq.~\eqref{eq:beta_effective} evaluates to $-\sigma^2 \beta_{\text{eff}}(t)$, where $\sigma$ is the width of the Gaussian. The time dependence follows trivially as $\beta_{\text{eff}}(t)=\beta_{\text{eff}}(0) \cos(\mu t) $. 

As a final remark, we notice that the perfect knowledge of the dynamics carried by Eq.~\eqref{eq:Heisenberg_H} and Eq.~\eqref{eq:Heisenberg_k} allows us to get an exact solution for the dynamics of a generic TFD $\ket{\beta}$ 
\be 
\ket{\beta(t)} = 
\frac{1}{\sqrt{Z(\beta)}} e^{-i \hat{n} t} e^{-\frac{\beta}{2} \HH} \ket{0} =  \frac{1}{\sqrt{Z(\beta)}}  e^{-\frac{\beta}{2} (\HH \cos(\mu t) - 2  \kk   \sin \mu t )} \ket{0} \label{eq:t_evolved_TFD}
\ee 
where we notice the relative minus sign from switching back to the Schr\"odinger picture. At $t=\pi/\mu$, $\ket{\beta} \rightarrow \ket{-\beta}$ and the return probability dips to $F(\pi/\mu) = (Z(0)/Z(\beta))^2$, which under mild assumptions can be seen to be the minimal value. 
\section{An approximate realization: chord states in SYK}
\label{sec:approx_realization_SYK}
As already emphasized in Section \ref{sec:intro}, the construction that we have presented in the previous Section relies on an implicit definition of the grading operator $\hat{n}_0$. The goal of the remaining of the paper is to present an explicit, yet approximate, realization of the operator $\hat{n}_0$ and of its extension $\hat{n}$, in the context of the SYK model.
To this end, we will start in section \ref{sec:SYK} by introducing a model made with two copies of the SYK model \cite{Maldacena_16, polchinski2016the} coupled via a simple bilinear term, as in Eq.~\eqref{eq:S_operator_def}, built in terms of the Majoranas across the two systems. 
We will show how the construction of Section \ref{sec:general_construction} can be carried over for the \textit{double-scaled} SYK (DSSYK) model, Eq.~\eqref{eq:H_tot_def}, in which the total number of fermions, $N$, and the number of fermions participating in each interaction term, $p$, are both sent to infinity keeping the ratio $p^2/N$ fixed. This particular choice has the advantage that the theory becomes analytically tractable. Another widely studied limit is obtained by sending $p$ to infinity only \textit{after} the $N \rightarrow \infty$ limit. This limit is known to qualitatively capture many of the features of the, more natural, fixed $p$ SYK model in the $N \rightarrow \infty$ limit. 

The three limits just described certainly give rise to distinct behaviors. However, we emphasize that this construction applies to all of them, with the degree of approximation scaling differently, and that the revival dynamics is clearly visible even in the simple case with both $N$ and $p$ fixed, thus showing the robustness of the analytical results obtained for the DSSYK case. 
\subsection{Two-Sided Double-Scaled SYK}
\label{sec:SYK}
Let us consider the case in which the two copies are  made of $N$ Majorana fermion operators each, \textit{i.e.} a set of $N$  operators (with $N$ even), named $\hat{\psi}_i^L$ and $\hat{\psi}_i^R$ ($i = 1,~\dots , ~N$), satisfying the following anti-commutation relations
\begin{equation}
\label{eq:anticommutation_relations}
    \left\{\hat{\psi}_i^L, ~\hat{\psi}_j^R  \right\} = 0 , \ \left\{\hat{\psi}_i^L, ~\hat{\psi}_j^L  \right\} = \left\{\hat{\psi}_i^R, ~\hat{\psi}_j^R  \right\} = 2 \delta_{ij} \, ,
\end{equation}
with $\delta_{ij}$ being the Kronecker delta. The system is described by the following Hamiltonian:
\begin{equation}
    \label{eq:H_tot_def}
    \hat{H}(\mu) \equiv \hat{H}_L - \hat{H}_R + \mu  \hat{S} \, ,
\end{equation}
where $\hat{H}_L$ denotes a SYK Hamiltonian \cite{Maldacena_16, polchinski2016the} with interaction of order $p$, acting on the left system. More explicitly, introducing the notations $I \equiv \left\{i_1,\dots, i_p\right\}$ (with the coefficients $i_j$ ordered and all different) and $\hat{\Psi}_I^L \equiv \psi_{i_1}^L \cdots \psi_{i_p}^L$, $\hat{H}_L$ reads:
\begin{equation}
    \label{eq:H_L_large_p_notation}
    \hat{H}_L = i^{p/2}\sum_{I} J_{I} \hat{\Psi}_I^L \, ,
\end{equation}
with the coupling constants $J_I$ being independent random Gaussian variables\footnote{In the double scaling limit, Gaussian distribution of $J_I$ is less relevant for solving the model, as long as $J_I$ are independent random variables with bounded moments. We consider Gaussian distributed $J_I$ to make the relation with other limits of SYK more transparent.} satisfying
\be\langle J_I \rangle_J = 0 ~~~~ \langle J_I^2 \rangle_J= \frac{\sJ^2}{\lambda \binom{N}{p}}  ~~~~~ \lambda \equiv (2p^2)/N  \ee 
where $\langle~,~\rangle_J $ denotes ensemble averaging and $\sJ$ is a constant setting the energy scale. 
$\hat{H}_R$ denotes a corresponding SYK Hamiltonian acting on the right system, with the disorder variables \textit{fully correlated} with their left counterparts \textit{i.e}
\begin{equation}
    \label{eq:H_R_def}
    \hat{H}_R = i^{p/2}\sum_{I} J_{I}^R \hat{\Psi}_I^R \, , \qquad J_I^R = i^p J_I \ \forall I \, .
\end{equation}
Finally, $\hat{S}$ denotes the following quadratic term
\begin{equation}
    \label{eq:S_operator_def}
    \hat{S} = \frac{1}{2} \sum^{N}_{j=1} ( 1+ i \,  \hat{\psi}^{L}_{j}\hat{\psi}^{R}_{j}  )=  \sum_{j = 1}^N \hat{c}^\dagger_j \hat{c}_j \, ,
\end{equation}
where we defined the fermionic annihilation operators $\hat{c}_j \equiv \frac{1}{2}(\hat{\psi}^{L}_{j}+i \hat{\psi}^{R}_{j})$. $\hat{S}$ has played an important role in the study of SYK, especially in the context of traversable wormholes \cite{maldacena2018eternaltraversablewormhole, Maldacena_2017}, and in microscopic realizations of how the $SL(2, R)$ generators of near-$AdS_{2}$ gravity are embedded in SYK \cite{Lin_2019, Lin_2023}. It is worth mentioning that, because of the different choice for the relative sign between the left and right Hamiltonians, the full Hamiltionian in Eq.~\eqref{eq:H_tot_def} \textit{does not} have the $S$ mod 4 symmetry of the Maldacena-Qi model \cite{Garcia_Garcia_2019}, while it only retains the $S$ mod $2$ symmetry shared with standard SYK \cite{garcia-garcia2016spectral}. Notice also that, thanks to the presence of a chiral symmetry operator which anticommutes with $\hat{S}$ \cite{Garcia_Garcia_2019}, $\ket{0}$ gets a partner, another maximally entangled state, $\ket{\bar{0}}$ which is annihilated by $\hat{c}^{\dagger}_j$. $\ket{\bar{0}}$ is an equally valid seed for the construction, but will not be important for the discussion that follows.  

It has been shown explicitly in \cite{Garcia_Garcia_2019} that the ground state of $\hat{S}$ (which is a positive operator by construction), denoted by $\ket{0}$, can be written as the infinite temperature TFD state of the SYK model, in agreement with the requirements of the construction presented in Section \ref{sec:general_construction}.
The eigenstates of $\hat{S}$, denoted by $\ket{s}$, are then labeled by the string describing the fermionic excitations above the vacuum $\ket{0}$. 
Let us notice that, as a result of the definition of the $\hat{c}_j$ operators in terms of the Majorana operators, the state $\ket{0}$ satisfies the property $\hat{\psi}_i^L \ket{0} = - i \hat{\psi}_i^R \ket{0}$.
In consequence, we see that the degenerate states $\ket{s}$ form an useful basis since it is possible to choose them such that $\ket{s, I} \propto \Psi^{s}_{I} \ket{0}$, where $\Psi^{s}_{I}$ is a Majorana string of size $s$ specified by the index set $I$. Thus, $\ket{s,I}$ can be mapped to operators with a well-defined \textit{operator size}, a fact extensively used in \cite{Qi_2019}.  

We will consider the so-called ``double-scaling limit'' of the model \cite{Cotler_2017, Berkooz_2018}. This is a limit in which both $N$ and $p$ are sent to infinity, in such a way that the parameter $\lambda$ remains finite, \textit{i.e.}
\begin{equation}
    \label{eq:double_scaling_limit}
    p\to \infty, \quad N \to \infty, \quad \mathrm{s.t.} \ \lambda\equiv \frac{2p^2}{N} \ \mathrm{finite} \, ~~~\qq \equiv e^{-\lambda} .
\end{equation}
The SYK model has been heavily studied in this limit, since it becomes analytically more tractable and displays an emergent algebra \cite{Berkooz_2018,Lin_2023}. We define the normalized trace such that $\tr(\mathbb I) = 1$. The energy variance reads $\tr(\hat{H}^2_L) = \frac{\sJ}{\lambda}$, showing that it is extensive when $p$ is finite, while becomes bounded in the double-scaling limit with $\lambda$ finite. The bounded spectrum of DSSYK plays an important technical role. In the limit $\lambda \rightarrow 0$, the spectrum spreads out to infinity and thus this limit needs to be taken carefully. A careful treatment of the $\lambda \rightarrow 0$ limit recovers the two-stage limit of SYK where $p$ is taken to be large after sending $N$ to infinity first \cite{Goel:2023svz, Lin_2023}.  

To make the paper more accessible and self-contained, we are now going to present some basic elements of the techniques used to analytically treat the SYK model in the double-scaling limit, and how the two-sided description naturally emerges in this context.
\subsection{Chord diagrams in the double scaling limit}
\label{sec:chord_diagrams_intro}
One of the most appealing features of the (single-copy) SYK model in the double-scaling limit is that, after disorder averaging, the model can be solved via an expansion over \textit{chord diagrams} for a large range of temperatures and time. To illustrate the logic, we consider the averaged partition function of, say, $\hat{H}_L$ 
\be  \expval{\tr( e^{-\beta \hat{H}_L} ) }_{J} = \sum^{\infty}_{k=0} \frac{(-\beta)^k}{k!} m_k~~~~~m_k \equiv \expval{ \tr({\hat{H}_L}^k) }_J, \label{eq:partition_function_moments} \ee 
where the nontrivial moments are
 \be m_{2k} = i^{k p}  \sum_{I_1,  I_2, I_3 \ldots I_{2k}}  \langle J_{I_1} J_{I_2} \ldots J_{I_{2k}}  \rangle_{J} \tr( \Psi_{I_1} \Psi_{I_2} \ldots  \Psi_{I_{2k}} )  ~, \label{eq:wick}  \ee 
and we dropped the subsystem labels since we are talking about a single-site system. It is immediate to verify that the algebra in Eq.~\eqref{eq:anticommutation_relations} is invariant under sign flips of each Majorana fermion. Therefore, any non-trivial product of the fermions is traceless, thus forcing the Majorana operators in Eq.~\eqref{eq:wick} to appear in pairs. Disorder averaging is done using Wick contractions among $J_I$, which pair up different index sets $\{ I_k \}$ appearing in the sum.  The $\Psi_{I}$ are moved past each other using the rules  $\Psi_{I_j} \Psi_{I_k} =(-1)^{I_{j} \cap I_k} \Psi_{I_j} \Psi_{I_k} $ until the fermion strings with the same index sets are next to each other, at which point the trace is easily evaluated to give $\text{tr}(\Psi^2_{I}) =\text{tr}(\mathbb{I}) = 1$. 

Summarizing, the evaluation of the partition function in Eq.~\eqref{eq:partition_function_moments} is now reduced to the problem of counting the intersections of the index sets $\{ I_i \}$ appearing in Eq.~\eqref{eq:wick}. 
This combinatorial problem gets enormously simplified in the double scaling limit, as the probability of having three or more index sets with nonzero intersections vanishes, allowing us to restrict our consideration to \textit{pairwise} intersections.
Moreover, the probability for two distinct $I_{j}$ and $I_{k}$ of having $|{I_{j} \cap I_k}|=m$ becomes Poisson distributed with mean value $\lambda/2$. Therefore, the phase picked up on commuting $\Psi_{I_j}$ with $\Psi_{I_k}$ can be replaced by its average value $\mathfrak{q}=e^{-\lambda}$ \cite{Cotler_2017, Berkooz_2018, Garc_a_Garc_a_2018}. 

Let us briefly review how the results previously stated are obtained. For $N \gg p$ (a condition that is valid in the double scaling limit as well), any single fermion flavor, $\hat{\psi}_i$, has a probability $\frac{p}{N}$ of belonging to one index set. The probability that a single fermion flavor belongs to a product of $k$ fermion strings $\Psi_{I}$ is therefore $(\frac{p}{N})^k$. The expected number of intersections, summing over all flavors, is consequently $N (\frac{p}{N})^k$, which vanishes for $k>2$ in the double scaling limit. 
The above argument can be made more explicit. To this end, we note that the probability of $m$ intersections between two fermion strings $I_1$ and $I_2$ with lengths $s_1$ and $s_2$ is given by 
\be \
\text{Pr}(|I_1 \cap I_2 |=m ) ~~=~~\frac{1}{\binom{N}{s_1} \binom{N}{s_2}} \frac{N!}{(s_1 -m)! m! (s_2 - m)! (N-s_1 -s_2 + m)!}  \label{eq:fermion_intersection_statistics} ~, 
\ee 
from which, the Poisson distribution of $m$ follows in the double scaling limit when $s_1=s_2=p$. The explicit expression appearing in Eq.~\eqref{eq:fermion_intersection_statistics} will prove to be useful when we consider one of the sizes $s_1$ finite.  

Given this premise, we arrive at the fundamental result that $m_{2k}$ in the double scaling limit is computed using chord diagrams organized by the number of chords and the intersections between them. The factors of $H$ are represented as $2k$ nodes arranged on a circle (which represents the cyclicity of the trace). The arrangement of the $H$ nodes reflects the ordering within the trace. Each chord represents a Wick contraction between a pair of $H$. The relative contribution of distinct diagrams comes from counting the number of chord intersections, which can be treated as independent. All in all, we get
\be 
m_{2k} = \Big( \frac{\sJ^2}{\lambda} \Big)^k \sum_{\text{diagrams with k chords} } \mathfrak{q}^{\text{number of intersections}} \,. 
\label{eq:moments_by_chords}
\ee 
\begin{figure}
    \centering
    \subfigure[]{\includegraphics[width=0.25\linewidth]{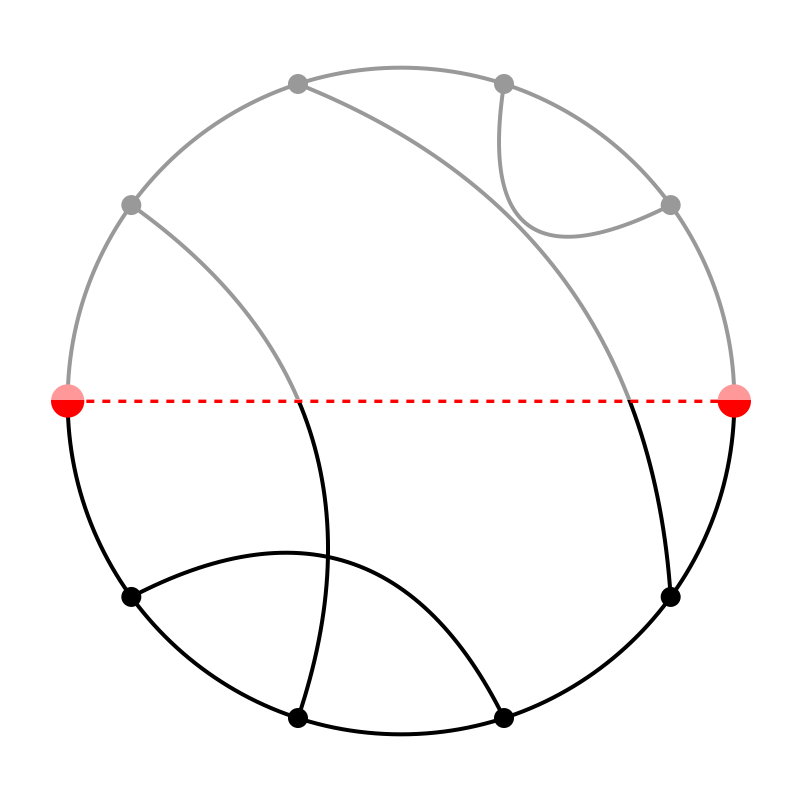}}
    \label{}
    \subfigure[]{\includegraphics[width=0.25\textwidth]{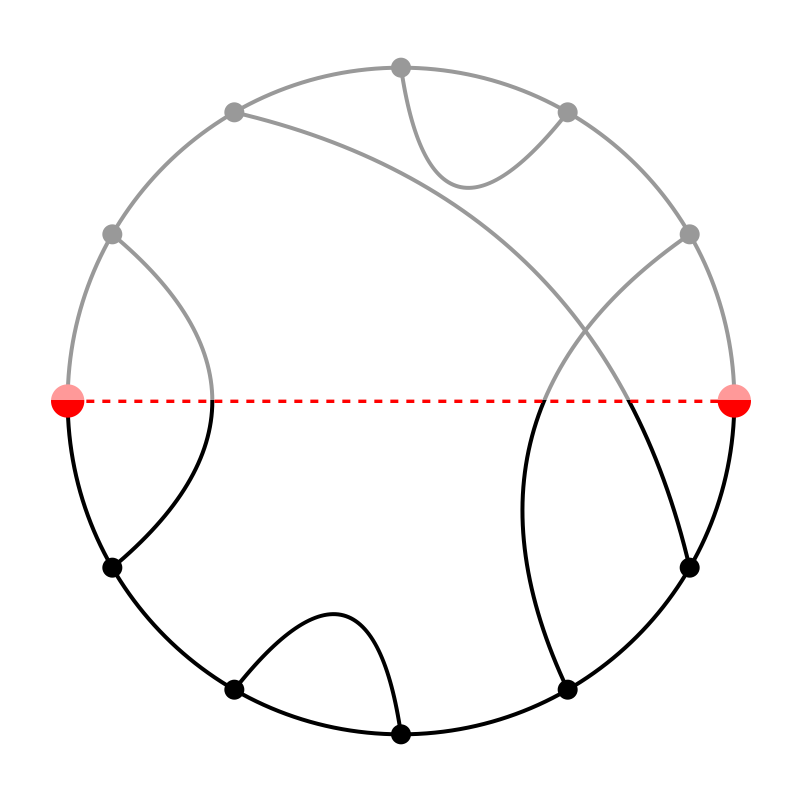}} 
    \captionsetup{justification=Justified}
    \caption{(a) The open chord state $\ket{2}$ from slicing open a chord diagram appearing in computation of $m_8$. (b) $\ket{3}$ contributing to $m_{10}$.  }
     \label{fig:chord_diagrams}
\end{figure}
The evaluation of Eq.~\eqref{eq:moments_by_chords} is further simplified by noticing that the process of adding further chords to a chord diagram has a recursive structure. This observation was used explicitly in \cite{Berkooz_2018}, and the problem of evaluating higher chord diagrams from lower ones was formulated in terms of a transfer matrix $\hat{T}$ on an auxiliary Hilbert space. The construction goes as follows.  A chord diagram is opened at a point (given the cyclicity of the problem, the point at which the cut is done is irrelevant), and the zero open chord state $\ket{0}$ -- which we will soon relate to the state $\ket{\beta=0}$ of the previous section -- is assigned to the open end. While moving along the circle, one interprets the emanation of new chords as a transition to a state with a higher ``open'' chord number $\ket{n}$ with $n>0$. The chord number decreases (i.e  $\ket{n} \rightarrow \ket{n-1} $) when both the nodes in a chord on the circle have been traversed. This process of traversing the chord diagram is weighted by the possibility of intersections between the different open chords. These rules are summarized as follows: 
\be
\hat{a} \ket{n} = \ket{n-1} ~~ \hat{a}^{\dagger} \ket{n} = \ket{n+1}~~~ \hat{n} \ket{n} = n \ket{n} \label{eq:chord_algebra} 
\ee
\be
\hat{T} = \frac{\sJ}{\sqrt{\lambda}}(\hat{a}^{\dagger} +  \hat{a} \frac{1-\qq^{\hat{n}}}{1-\qq}), ~~ m_{2k} = \bra{0} \hat{T}^{2k} \ket{0} ~,  
\label{eq:transfer_matrix_moments} 
\ee 
with $\hat{a}^\dagger$ ($\hat{a}$) being the creation (annihilation) operator for a chord state, and $\hat{n}$ being the operator that counts the number of open chords. 

The algebra satisfied by $\hat{a}$ and $\hat{a}^\dagger$ is more transparently stated in the \textit{normalized} chord basis $ \{ \ket{\nn} \}$, in which $\hat{T}$ becomes hermitian and reads
\be  
\hat{T} = \frac{\sJ}{\sqrt{\lambda}}( \aa + \aa^{\dagger} ) ~,
\ee 
while $\aa$ and $\aa^{\dagger}$ satisfy the following $\qq$-deformed algebra (see Appendix \ref{sec:Chord_Basis_Details} for more details)
\be  
\aa \aa^{\dagger} -  \qq \aa^{\dagger} \aa = 1. 
\label{eq:q_deformed_algebra}
\ee 
\subsection{A physical interpretation of the chord Hilbert space}
\label{sec:double_sided_geometry_chord_diagrams}
Up to here, the chord Hilbert space, the operators $\hat{n}$, $\hat{a}$, $\hat{a}^\dagger$, and the $\qq$-deformed algebra that they satisfy, have been introduced just as a very convenient computational tool. In particular, no clear connection with the physical Hilbert space of the DSSYK model has been provided.
This aspect has been investigated extensively in \cite{Lin_2022}, which (among other things) showed that the chord Hilbert space is naturally realized when \textit{two} copies of the SYK are considered together.

The technical reason behind this identification is easy to understand: since we are evaluating traces in the single-site DSSYK model, they can be equivalently evaluated as inner products in terms of the infinite-temperature thermofield double state $\ket{0}$, defined on the doubled Hilbert space. More generally, we have the following map between two-sided amplitudes and one-sided traces:
\be 
\bra{0} \hat{M}_{L} \hat{O}^1_{L} \hat{O}^{2}_{R} \hat{N}_{L}  \ket{0} \propto  \tr( \hat{M} \hat{O}^1 \hat{N} \hat{O}^{2}) ~, 
\label{eq:2copy_to_1copy_map} 
\ee
where $\ket{0}$ is the $\ket{\beta=0}$ state, the proportionality constant depends on the fermion parity of the operators, and we made use of the property $\hat{\psi}_i^L \ket{0} = - i \hat{\psi}_i^R \ket{0}$. A familiar version of this formula arises when we take $\hat{M}=\hat{N}= e^{-\beta \hat{H}}$, and Eq.~\eqref{eq:2copy_to_1copy_map} reduces to the familiar statement that expectation values in a TFD state reduce to one-sided (thermal) traces. 
Therefore, we see that the construction of the chord states takes the trace over one copy, as in the RHS of Eq.~\eqref{eq:2copy_to_1copy_map}, and maps it to the inner product of a corresponding two-sided chord state. $\hat{M}_{L}$ and $\hat{N}_L$ are used to do the state preparation, and $\hat{O}^1_{L} \hat{O}^{2}_{R}$ are operator insertions  
This process can be visually represented as in Fig.~\ref{fig:chord_diagrams}, where the chord diagram has been divided into two halves, with the bottom half below the red line representing the ket $\ket{n}$. The chord diagram is nonvanishing only if all the open chords leaving the bottom half are paired up with the ones in the top half. This means that states with different chord numbers are orthogonal. The intersections between the open chords from the ket state and the bra state are accounted for in the inner product. The same chord diagram can be sliced in different ways, which represents the freedom in writing $m_{2k}$ as a sum over inner-products of open chord states. 
Using that the infinite temperature TFD, $\ket{0}$, is the zero chord state, the connection with the moments given in Eq.~\eqref{eq:transfer_matrix_moments}, and that $\hat{T}$ is tridiagonal in the chord basis, it follows that $\hat{T}$ is identified with $\HH$, and the chord states span $\Hkryl$. This was already noticed in \cite{Lin_2022} and further studied in \cite{Rabinovici:2023yex, Ambrosini:2024sre}.
From Eq.~\eqref{eq:2copy_to_1copy_map}, we note that $\ket{0}$ provides an embedding of the empty chord state into the two-sided physical Hilbert space. Assuming self-averaging for the moments, $\HH$ written in the Krylov basis coincides with the transfer matrix $\hat{T}$. Thus the Krylov basis introduced in \eqref{eq:n_0_krylov_explicit}, for each disorder realization, gives an approximate embedding of the chord Hilbert space on the two-sided physical Hilbert space. This was already noticed in \cite{Lin_2022} and further studied in \cite{Rabinovici:2023yex, Ambrosini:2024sre, Xu:2024gfm, Balasubramaniam_2026, Nandy_2025}. 

The unnormalized states $\ket{n}$ are convenient for the combinatorics of the chord diagrams, but from now on we will use the normalized open chord states $\ket{\nn}$, see Appendix \ref{sec:Chord_Basis_Details} for the relation between them. 
Having identified that the chord diagram construction is naturally embedded in the two-sided SYK model, \cite{Lin_2022} also made the important observation that $\hat{n}$ has an approximate microscopic realization in terms of the operators $\hat{S}$ appearing in Eq.~\eqref{eq:S_operator_def}. To this end, we notice that, as long as $N \gg p$, the intersection of index sets among the chords is highly suppressed. In the double scaling limit, it is easy to quantify the variance of the operator $\hat{S}$ on the open chord states $\ket{\nn}$, \textit{i.e.} $\bra{\nn} \hat{S}^2 \ket{\nn} -\bra{\nn} \hat{S} \ket{\nn}^2$, using the Poisson distribution of intersections. Using the Poisson approximation (see Appendix \ref{sec:Chords_as_approx_eigenstates}) in the double scaling limit, we find
\be 
\bra{\nn}\hat{S} \ket{\nn} \approx n p -\Lambda_n, ~~~~   \bra{\nn} \hat{S}^2 \ket{\nn} -\bra{\nn} \hat{S} \ket{\nn}^2 \approx \Lambda_n ~~~ \Lambda_n = \binom{n}{2} \lambda.
\label{eq:size_variance_and_mean}
\ee 
Therefore, in the double scaling limit, we arrive at
\be 
\hat{n} \approx \frac{1}{p}\hat{S} ~,  \label{eq:approx_chord_number_op}
\ee 
where this operator equation is understood to hold with $O(\lambda/p)$ corrections, when restricted to $\Hkryl$. Thus Eq.~\eqref{eq:H_tot_def} provides a concrete example of Eq.~\eqref{eq:total_hamiltonian_mu-shifted} in the SYK model, with corrections that vanish in the double scaling limit. 

Before concluding this overview, we notice that \cite{Lin_2022} set up a precise dictionary between the chord states described above and the Hilbert space of JT gravity without matter.  $\HH$ in the  $\{ \ket{\nn} \}$ basis becomes the Liouville Hamiltonian \cite{Bagrets_2016, Harlow_2019} in the limit $\lambda \rightarrow 0$, $n \rightarrow \infty$, with $l= \lambda n + 2 \log(\lambda) $ held fixed.  
The dictionary just outlined can also be extended to the richer case of JT gravity with matter. On the DSSYK side, this boils down to the insertion of ``matter'' chords corresponding to random operators in addition to the Hamiltonian chords described above. These random operators are of the same form as in Eq.~\eqref{eq:H_R_def} with $p$ replaced by $r$ with $r^2/N$ held finite. $(H_L - H_R)$ does not annihilate the states with nontrivial matter chord insertions, and we will not directly use them in the remainder of this paper. 

\subsection{Quantum many-body scars in the two-sided model: regimes of validity}
\label{sec:validity_of_scars}
The discussion above clarifies that the full Hamiltonian of Eq.~\eqref{eq:H_tot_def} realizes, in the double-scaling limit, an \textit{approximate} yet explicit realization of the many-body scars construction described in Section \ref{sec:general_construction}. On the one hand, this realization has the merit of being explicit and based on the operator $\hat{S}$, which can be concisely written in terms of the microscopic operators $\hat{\psi}_i^L$ and $\hat{\psi}_i^R$. On the other hand, the construction is intrinsically approximate in nature, even in the double-scaling limit. In the remainder of this section, we will present a general analysis of the robustness of the revival dynamics, both within and outside the strict double-scaling limit. 

Note that $\hat{S}$ \textit{is not} an endomorphism of the chord Hilbert space. Therefore, it is clear that the dynamics induced by the Hamiltonian appearing in Eq.~\eqref{eq:H_tot_def} will only approximately reproduce the non-ergodic dynamics discussed in Section \ref{sec:general_construction}, and a certain amount of ``leakage'' out of $\hat{H}_\mathrm{Krylov}$ will be inevitable. Therefore, it becomes important to get an estimate of how strong the effects of this approximate dynamics are on states living in $\hat{H}_\mathrm{Krylov}$. Looking at Eq.~\eqref{eq:size_variance_and_mean}, it is clear that in the double-scaling limit, the only states for which $\hat{S}$ will vary substantially from $\hat{n}$ are the states $\ket{\nn}$ with $n$ being divergent. On the other hand, whenever $p$ is finite, the property of $\ket{\nn}$ being an approximate eigenstate of $\hat{S}$ deteriorates while increasing the (finite) value of $n$. Given that the chord Hilbert space is infinite, it becomes important to treat carefully states with overlap on chord states with divergent $\Lambda_n$, to understand the extent to which they will realistically show a non-trivial revival dynamics. For the non-ergodic dynamics described in Section \ref{sec:general_non_ergodic_dynamics}, the return amplitude defined in Eq.~\eqref{eq:survival_amplitude} is a basic probe. In this setup, let us introduce a time-scale, which we will call $\td$, quantifying the extent of time around $t=0$, for which Eq.~\eqref{eq:survival_amplitude} is a good approximation of the actual dynamics on $\Hkryl$. On general grounds (see Appendix \ref{sec:Chords_as_approx_eigenstates} for more details) we can write
\be 
\td \sim \frac{1}{\mu \sqrt{\lambda \expval{n^2}}},~~~~~~\expval{n^2} = \sum_{n} \abs{c_n}^2 n^2, ~~~~~~ c_n \equiv \bra{\nn}\ket{\psi_0} ~, 
\label{eq:tdep_def}
\ee
where the initial state has been expanded in the chord basis.
Taking $c_n=\delta_{mn}$, we get the time scale during which the chord state $\ket{\textbf{m}}$ acts like an approximate eigenstate for $\hat{H}(\mu)$ with eigenvalue $ \sim \mu m p $. We are discussing a limit where the spectrum of a generic Hamiltonian is continuous, so an approximate eigenstate means a wavepacket with narrow energy distribution, which evolves coherently until a time scale inverse to the energy width. This means that the results of Section \ref{sec:general_construction} can be imported to the approximate setting for times up to order $\td$. 
Given Eq.~\eqref{eq:survival_amplitude}, we immediately see that $\fid(t)$ shows revivals with a period 
\be 
\label{eq:t_rev}
\trev \approx \frac{2 \pi}{ \mu p} ~, 
\ee 
From which we see that the system will show \textit{nearly complete} revivals for the order of $\trev/\td \sim \frac{p}{\sqrt{\lambda}} = \sqrt{N}$ cycles, assuming that $\expval{n^2}$ is finite. 

Let us now overview the revival properties for several sets of initial states, and we refer to Section \ref{sec:dynamics_double_scaling} for more details. In the strict double-scaling limit, with finite $\lambda>0$, generic TFDs $\ket{\beta}$, having $\beta$ not scaling with $p$, will have finite values of $\expval{n^2}$.
Therefore, $\mu$ can be always chosen in such a way that $\td$ is as large as desired. For the finite $\lambda$ case that we discuss in the next Section, we will always consider states for which the corrections due to $\Lambda_n$ can be ignored, and therefore we will have perfectly persistent revivals. Setting $\mu=\frac{\omega}{p}$ with $\omega$ being $O(1)$, makes the revivals to happen on a finite time-scale $2 \pi/\omega$. 
Therefore, we conclude that the TFD states $\ket{\beta}$ form a well-characterized set of initial states for which the revival dynamics will be clearly visible, at least in the double-scaling limit.

In the double-scaling limit, the Arik-Coon $\qq$ coherent states are another class of states showing an intriguing dynamics. These states have a time-independent uncertainty for $\HH$, and in some sense represent a classical orbit for $\HH$ and its conjugate $\hat{k}$. The projection to $\HH$ describes how the particle moves on the single-site spectrum, and for these states we see that the particle moves with a fixed shape. Remarkably, the uncertainty in the $\qq$ coherent states measures the value of the Liouville potential evaluated on-shell. A related property is that for $\qq$ coherent states, the uncertainty in $\HH$ can be made arbitrarily small, which implies that two-sided DSSYK eigenstates emerge as a limit of the coherent states. In this case, the coherent state dynamics describes a sharply localized particle on the spectrum, moving through the eigenstates.  
The $\lambda \rightarrow 0$ case needs special treatment. The corrections due to $\Lambda_n$ are favorably suppressed in this limit. However, in this limit, $\ket{\beta}$ has size diverging with $N$  \cite{Qi_2019, Lin_2023, Okuyama_2023} though with fluctuations small compared to the mean. 
For such states, we do not get revival dynamics but a qualitatively different kind of non-ergodic dynamics that amounts to taking a continuum limit as discussed in Section \ref{sec:continuum_limit}. This continuum limit is very closely related to the identification of the length operator in JT gravity as a continuous counterpart of the discrete chord number operator: $\lambda \rightarrow 0$ such that the  length $l=\lambda n + 2 \log(\lambda)$ is finite.  

Applying Eq.~\eqref{eq:Heisenberg_EOM} with the identification of $\hat{n} \approx \hat{S}/p$ leads to an interesting alternative perspective on the revival dynamics. $\hat{S}$ is a Hamiltonian quadratic in $c_i$ and so we obtain
\begin{align}
    e^{i \hat{S} \omega t} \hat{c}_i e^{-i \hat{S} \omega t} & = \hat{c}_i e^{-i  \omega t }.  \label{eq:fermion_oscillator_complex} \\ 
  e^{i \hat{n} \omega t} \hat{\psi}^{i}_{L} e^{-i \hat{n} \omega t}  &=  \cos(\frac{\omega}{p} t) \hat{\psi}^{i}_{L} + i \sin(\frac{\omega}{p} t) \hat{\psi}^{i}_R   \label{eq:fermion_oscillator_majorana}
\end{align}
thus the dynamics is that of decoupled fermionic oscillators in this approximation. The oscillations are in the two-level space spanned by the occupation number basis of $\hat{c}_{i}$, delocalized across the Majorana $\hat{\psi}^{i}_{L}$ and its partner in the right system. While the fermions oscillate independently, their motion between the left and right systems is correlated. Therefore despite each individual fermion oscillating very slowly, a fermionic string of $O(p)$ weight can have these phases aligned to have collective motion on finite time scales. This is what gives rise to the classical motion of the operator $\HH$ in Eq.~\eqref{eq:Heisenberg_H}. The fact that the right hand side of Eq.~\eqref{eq:Heisenberg_H} takes this precise form is a nontrivial consequence of the DSSYK rules. The fermion strings in $\HH$ that would oscillate with a frequency smaller than $\omega$, say, $2\omega/3, \omega/2, \omega/3/$ have vanishing matrix elements in the chord basis. 

We can reinterpret the reflection on the spectrum $\langle \HH(t=\frac{\pi}{\omega}) \rangle = - \langle \HH(0) \rangle $ in light of  Eq.~\eqref{eq:fermion_oscillator_majorana}. At $t=\frac{\pi}{\omega}$, the phases of fermions interfere constructively to give a microscopic explanation for $\hat{H}_{L} \rightarrow -\hat{H}_{L}$ in the subspace and the same for $\hat{H}_{R}$. This also helps us see why $A(t=\frac{\pi}{\omega}) = Z(\beta)^{-1}$ for the thermofield double $\ket{\beta}$.    
 
\section{Dynamics in the double-scaling limit}
\label{sec:dynamics_double_scaling}
In this section, we discuss in more details the non-ergodic dynamics induced by the Hamiltonian of Eq.~\eqref{eq:H_tot_def} on the Krylov subspace spanned by the chord states $\ket{\nn}$. This is a specific realization of the construction presented in Section \ref{sec:general_construction}, with the interaction $\hat{S}$ providing an approximate lift of the operator $\hat{n}_0$. The details of the approximation  are summarized in Eq.~\eqref{eq:size_variance_and_mean}, and its consequences on the dynamics of the chord states is addressed in Eq.~\eqref{eq:tdep_def}. In Section \ref{sec:revival_dynamics_DSSYK}, we discuss states for which $\td$ (and the ratio $\td/t_{\text{rev}}$) is taken to infinity.

\subsection{Revivals in DSSYK}
\label{sec:revival_dynamics_DSSYK}
The general features of the scar dynamics discussed in Section \ref{sec:general_non_ergodic_dynamics} carry through here. On the other hand, there are more properties, which are dependent on the specific realization of $\hat{L}_{\pm}$, that we will discuss explicitly here. The crucial detail is the representation of the operators $\hat{L}_{\pm}$ in the chord basis, from which we can also determine the commutation relation between them. The chord creation $\aa^{\dagger}$ and annihilation $\aa$ operators play the role of $\hat{L}_{+}$ and $\hat{L}_{-}$, respectively, and they satisfy the $\qq$ deformed algebra of Eq.~\eqref{eq:q_deformed_algebra}. The model is solved using $\qq$-deformed functions, and we relegate the technical details of the solution to Appendix \ref{sec:Chord_Basis_Details}. We will make use of the $\qq$ Pochammer symbol $(a, \qq)_n$ and the $\qq$ integer $[n]_{\qq}$ in this Section, which are defined as follows:
\be (a, \qq)_n \equiv \prod^{i=n-1}_{i=0} (1- a \qq^{i})~~~ [n]_{\qq} \equiv \frac{1-\qq^n}{1-\qq}.  \ee 

Let us recall that, when acting on $\mathcal{H}_\mathrm{Krylov}$, The Hamiltonian $\HH$ is identified with the transfer matrix $\hat{T}$ and therefore, when expressed in the $\ket{\nn}$ basis,  it takes the tridiagonal form
\be 
\HH =\frac{\sJ}{\sqrt{\lambda}} (\aa+\aa^{\dagger}), \label{eq:normalized_chord_hamiltonian} 
\ee 
from which, the eigenpairs read
\be 
\HH \ket{\theta} = E(\theta) \ket{\theta}, ~~ E(\theta)= \frac{2 \sJ \cos(\theta) }{ \sqrt{\lambda(1-\qq)} } \label{eq:spectrum_of_DSSYK} ~~ 
\ee 
where the angle $\theta \in [0, \pi]$ parametrizes the eigenstates. The spectrum of Eq.~\eqref{eq:spectrum_of_DSSYK} is obtained from the behavior of $\hat{T}$ at asymptotically large index $n$ \cite{Berkooz_2018}. 
From the very definition of $\mathcal{H}_{\mathrm{Krylov}}$, it follows that the energy eigenstate $\ket{\theta}$ represents a tensor product of the SYK eigenstates on the left and right system with same energy, \textit{i.e.} $\ket{E(\theta)}_{L} \ket{E(\theta)}_{R}$. The spectrum is continuous, with $E(\theta)$ taking values in  $[ \frac{- 2\sJ  }{ \sqrt{\lambda(1-\qq)} },  \frac{2\sJ  }{ \sqrt{\lambda(1-\qq)} }]$. The eigenstates are $\delta$ normalized as $\bra{\theta_1} \ket{\theta_2} = \delta(\theta_1- \theta_2 )/\rho(\theta_1)$, where $\rho(\theta)$ is the density of states, see Eq.~\eqref{eq:rho(theta)} for the explicit form. 

The presence of a continuous energy spectrum makes the current setup slightly different from what is discussed in Section~\ref{sec:general_non_ergodic_dynamics}, but it is a natural consequence of having already taken the $N\rightarrow \infty$. The bandwidth of the spectrum is encoded in the $\lambda$ dependence, and when the $\lambda \rightarrow 0$ limit is taken, the maximum energy scales extensively with $N$. For the remainder of this section, we set $\sJ=-\sqrt{\lambda}$.  ~\

The possibility of finding the eigenpairs of $\HH$, Eq.~\eqref{eq:spectrum_of_DSSYK}, implies that the basis $\{ \ket{\theta} \}$ is known in terms of the chord basis $\{ \ket{\nn} \}$ and vice-versa. The coefficients $\phi_n(\theta) = \bra{\theta} \ket{\nn}$ are given by the continuous $\qq$-deformed Hermite polynomials (see Eq.~\eqref{eq:qHermite}). Therefore, the time-evolved state $\ket{\psi(t)}$ can be expanded both in the chord basis, see Eq.~\eqref{eq:amplitude_DS_limit}, and in the energy basis, $\ket{\theta}$. Following the latter choice, we expand $\ket{\psi(t)}$ as 
\be 
\ket{\psi(t)} = \int^{\pi}_{0}  f_{\psi}(\theta, t) \rho(\theta)  \ket{\theta} d\theta,  
\ee 
with $f_{\psi}(\theta, t) = \bra{\theta } \ket{\psi(t)}$. This is the profile of the time-evolved state in the energy basis, the analog of $c_k(t)$ in Eq.~\eqref{eq:Psi_decomposition_LR_energy_basis}, except that it is parametrized by the continuous spectral angle $\theta$. Given this parametrization, we can write the overlap 
\be A(t_1 - t_2) =
\bra{\psi(t_1)} \ket{\psi(t_2)} = \int^{\pi}_{0} f^{*}_{\psi}(\theta, t_1) f_{\psi}(\theta, t_2) \rho(\theta) d \theta \implies  \int^{\pi}_{0} \abs{f_{\psi}(\theta, t)}^2 \rho(\theta) d \theta  = 1, \label{eq:survival_amplitude_as_f_convolution}
\ee 
which provides a concrete way to visualize the motion of the particle on the spectrum as discussed in Sec.~\ref{sec:general_non_ergodic_dynamics}. To explicitly compute $f_{\psi}(\theta, t)$, we can introduce the propagator $K(\theta, \theta^{\prime}, t) = \bra{\theta} e^{-i \hat{n} \omega t} \ket{\theta^{\prime}}$, which is the Poisson kernel of the $\qq$ Hermite polynomials, see Eq.~\eqref{eq:generating_function_series}. Given $K(\theta, \theta^{\prime}, t)$, the function $f_{\psi}(\theta, t)$ is determined from the initial conditions $f_{\psi}(\theta^{\prime}, 0)$ as follows  
\be 
f_{\psi} (\theta, t) =  \int^{\pi}_{0}  K(\theta^{\prime}, \theta, t)   f_{\psi}(\theta^{\prime}, 0) \rho(\theta^{\prime}) d \theta^{\prime}.  \label{eq:reproducing_Poisson_kernel}
\ee 
The representation Eq.~\eqref{eq:reproducing_Poisson_kernel} becomes particularly useful when considering a TFD state $\ket{\beta}$, defined through initial data $f_{\beta}(\theta, 0)=  e^{-\frac{\beta}{2} E(\theta)}/\sqrt{Z(\beta)}$. 
The kernel $K(\theta^{\prime}, \theta, t)$ is proportional to $\delta(\theta^{\prime}- \theta)$ at $t=0$. From Eq.~\eqref{eq:reproducing_Poisson_kernel}, using that the angular variable appears in pairs of both signs, we deduce the reflection conditions 
\be 
\text{Re}  f_{\psi} (\theta, t) = \text{Re} f_{\psi} (\pi-\theta, \pi \pm \omega t), ~~~ \text{Im}  f_{\psi} (\theta, t) = \mp \text{Im} f_{\psi} (\theta, \pi \pm \omega t).
\label{eq:reflection_conditions}
\ee 
This is obtained from the explicit form of Eq.~\eqref{eq:reproducing_Poisson_kernel} and illustrates how the wavefunction bounces back and forth in time, satisfying Eqs.~\eqref{eq:Heisenberg_H} and \eqref{eq:Heisenberg_k}. The divergence of the kernel as $\theta^{'} = \theta+\omega t$, reflects that eigenstates $\ket{\theta}$ undergo sharply localized motion, which is \textit{ballistic} for the spectral angle. We will reinterpret this in the light of $\qq$ coherent state dynamics in Sec~\ref{sec:coherent_states}.  
The expansion of $\ket{\beta}$ in the chord basis was computed in \cite{Okuyama_2023} and is presented in Eq.~\eqref{eq:tfd_in_chord_basis}. Although its expression is not particularly concise, it proves to be quite convenient for numerically evaluating simple expressions in the chord basis, including a numerical evaluation of quantities like $A(t)$, $\HH(t)$ and $\kk(t)$. It turns out to be especially convenient for small enough $\beta$ and $\qq$, when the infinite sums in both $r$ and $n$ (see Eq.~\eqref{eq:tfd_in_chord_basis}) can be truncated to sufficiently few terms. 

\begin{figure}[t!]
    \centering
    \includegraphics[width=0.98\linewidth]{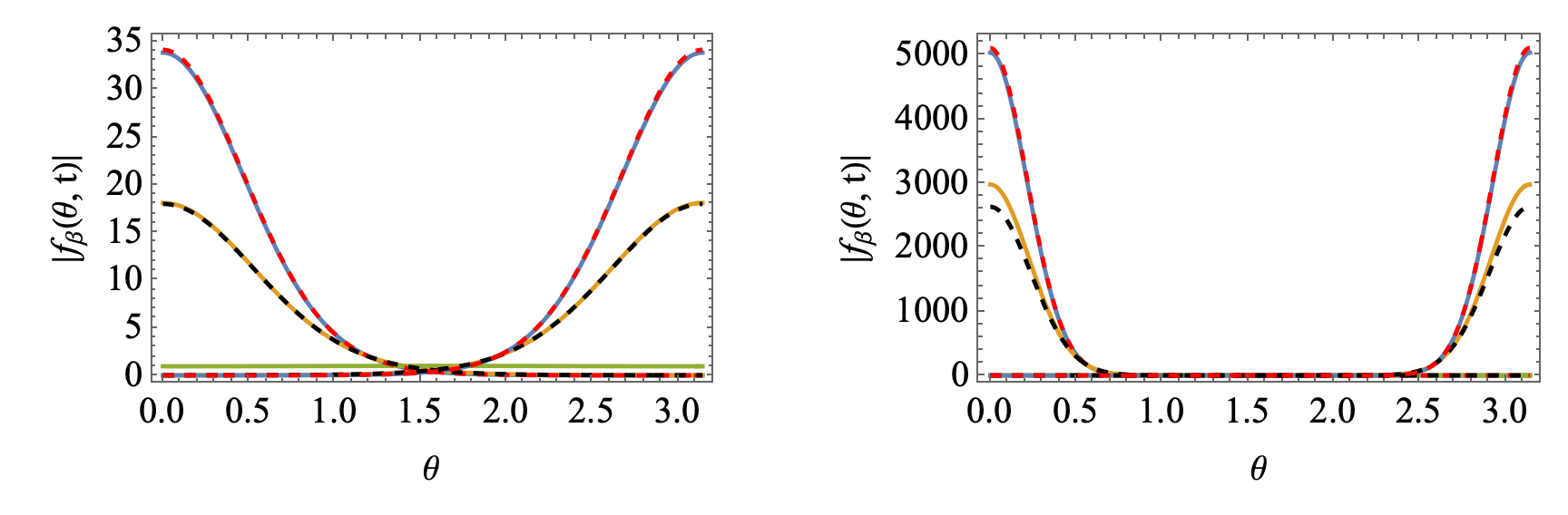}
    \captionsetup{justification=Justified}
    \caption{
$\abs{f_{\beta}(\theta, t)}$ for $\qq=0.8$, $\beta=2$ (left) and $\beta=10$ (right). For $\beta=2$, we sample $\abs{f_{\beta}(\theta, t)}$ for $\frac{\omega t}{2 \pi} = 0$ (blue),  $\frac{\omega t}{2 \pi} = 0.1$ (orange), and $\frac{\omega t}{2 \pi} = 0.25$ (green). For $\beta=10$, we show $\frac{\omega t}{2 \pi} = 0$ (blue),  $\frac{\omega t}{2 \pi} = 0.05$ (orange), and $\frac{\omega t}{2 \pi} = 0.25$ (green). The distribution gets reflected around $\theta=\frac{\pi}{2}$ at $\pi -\omega t$, we have plotted reflected distributions for the specified times with the same colors. The dashed lines are the result of fitting with the Gibbs form $\sim e^{-\frac{\beta_{\text{eff}}(t)}{2} E(\theta)}$ with $\beta_{\text{eff}}(t)$ given by Eq.~\eqref{eq:beta_effective}. The agreement is approximate in nature. The width of $\abs{f_{\beta}(\theta, t)}$ changes as a function of $t$ to satisfy Eq.~\eqref{eq:survival_amplitude_as_f_convolution}.   }
    \label{fig:Abs_f}
\end{figure}

\begin{figure}[t!]
    \centering
    \includegraphics[width=0.9\linewidth]{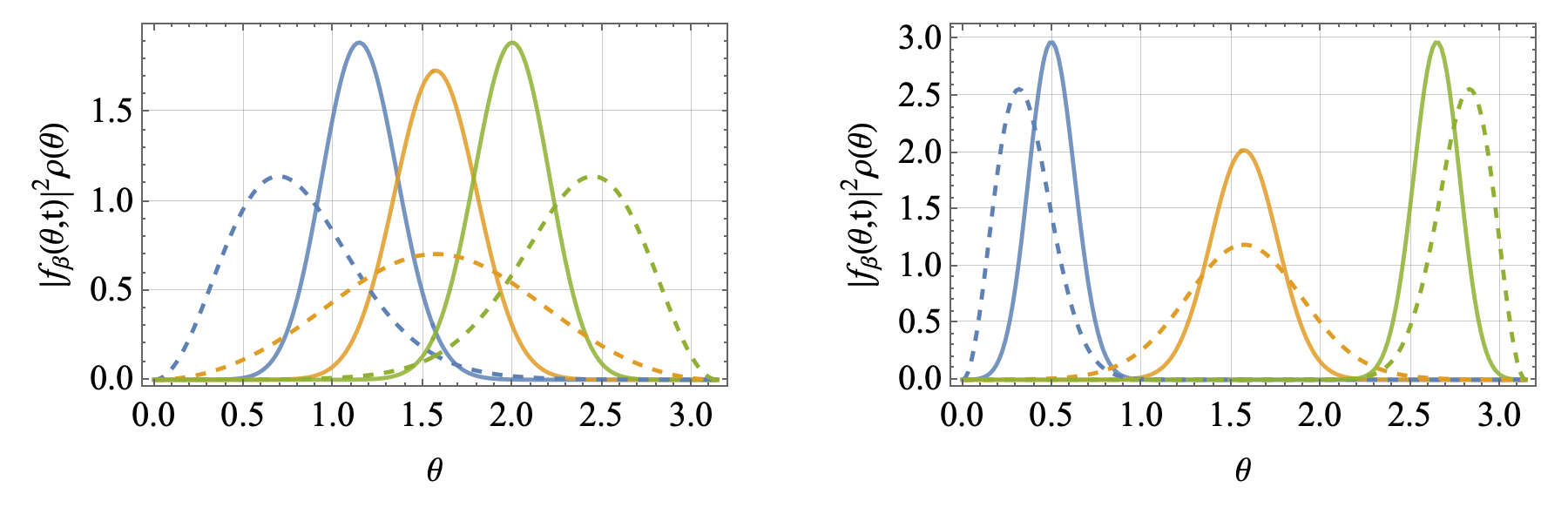}
    \captionsetup{justification=Justified}
    \caption{ The normalized wavepacket profile $\abs{f_{\beta}(\theta, t)}^2 \rho(\theta)$  for $\beta=2$ (left) $\beta=10$ (right) with $ \qq=0.05$ (dashed lines) and $ \qq=0.8$ (solid lines). We show the densities moving from the left at $t=0$ (blue) to a configuration symmetric about $\theta=\frac{\pi}{2}$ at $t=\frac{\pi}{2 \omega}$ (orange) to $t=\frac{\pi}{\omega}$ (green), when it gets reflected about $\theta=\frac{\pi}{2}$. For fixed $\beta$, smaller $\qq$ gives a broader distribution in $\ket{\theta}$ which can be traced back to how $\rho(\theta)$ depends on $\qq$. On increasing $\beta$, the wavepacket becomes more sharply peaked for the entire cycle. }
    \label{fig:rho_times_Abs_f}
\end{figure}

\begin{figure}
    \centering
    \includegraphics[width=0.99\linewidth]{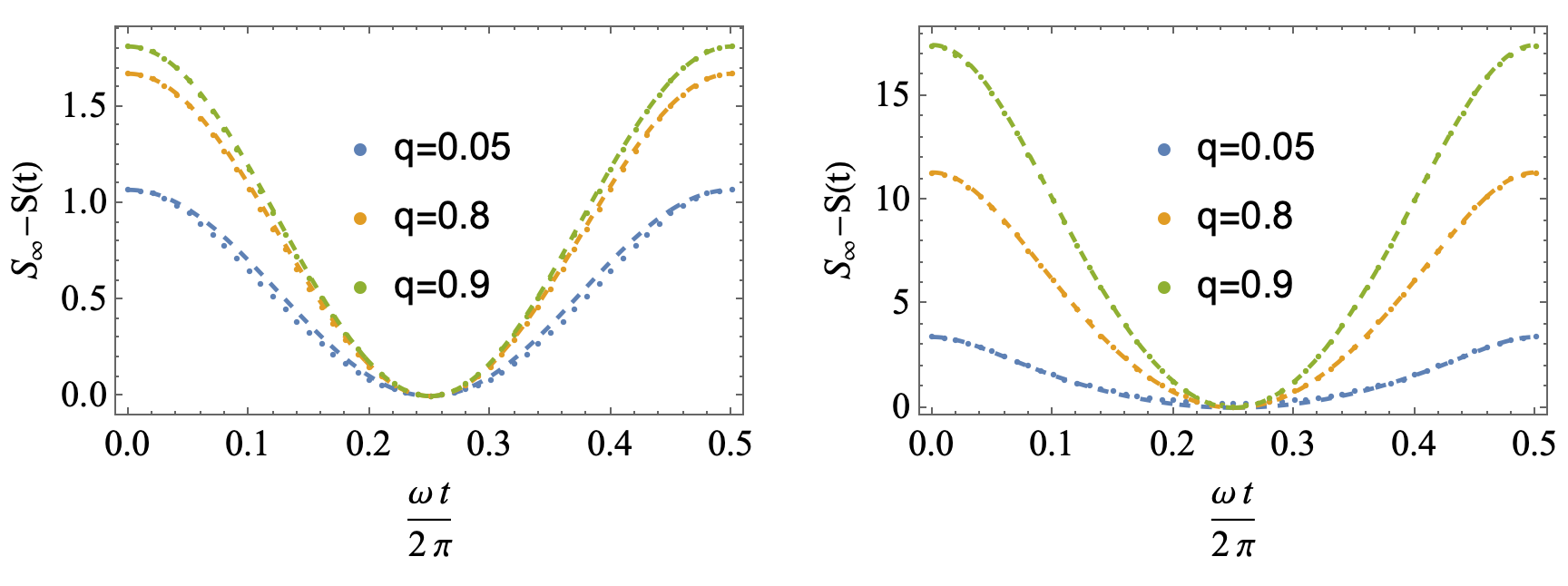}
    \captionsetup{justification=Justified}
   \caption{The entanglement entropy difference $S_{\infty}-S(t)$ for $\ket{\beta(t)}$ with $\beta=2$ (left) and $\beta=10$ (right). $S(t)$ is the von Neumann entropy of $\rho_{L/R}(t)$ as defined in  Eq.~\eqref{eq:one-sided}. The entanglement entropy difference is correlated with $\langle \HH(t) \rangle $ and $\langle \kk(t) \rangle$. The entanglement entropy is well approximated by the thermal entropy evaluated at $\beta_{\text{eff}}(t)$ (dashed lines).  For $\beta=2$, $E(\beta) \approx \beta$ across different values of $\qq$ and $\beta_{\text{eff}}(t) \approx \beta(0) \cos(\omega t)$. We used the Gaussian form $S_{\infty}-S(t) \propto \beta(0)^2 \cos^2(\omega t)$ in the left panel as $S(\beta_{\text{eff}}(t))$ and found good agreement. For $\beta=10$, we computed $\beta_{\text{eff}}(t)$ by directly numerically solving for $\beta_{\text{eff}}(t)$ in Eq.~\eqref{eq:beta_effective}}
    \label{fig:Entanglement_Entropy_DSSYK}
\end{figure}

The function $f_{\beta}(\theta, t)$ can be directly connected to observables.
We show examples of the absolute value $\abs{f_{\beta}(\theta, t)}$ in Fig.~\ref{fig:Abs_f}. All the one-sided expectation values for the left (right) system are determined by the combination $|f_{\beta}(\theta, t)|^2 \rho(\theta)$, see Eq.~\eqref{eq:one-sided}. The profile $f_{\beta}(\theta, t)$ captures the motion of the normalized wavepacket, for which examples are presented in Fig.~\ref{fig:rho_times_Abs_f}. 
A quantity which concisely captures the form of the distribution  $|f_{\beta}(\theta, t)|^2 \rho(\theta)$ is the total entanglement entropy between the left and the right system $S(t)$. In the double-scaling limit, $S(t)$ is divergent, but the entropy difference $S_{\infty}-S(t)$ from the maximally entangled state, $S_{\infty}$, is well-defined. Fig.~\ref{fig:Entanglement_Entropy_DSSYK} features the entropy difference, and it shows a remarkable agreement with the thermal entropy difference, with the effective temperature determined by the value of $\langle \HH(t) \rangle$ as in Eq.~\eqref{eq:beta_effective}. 

The agreement between $\abs{f_{\beta}(\theta, t)}$ and the thermal form  $e^{- \frac{1}{2} \beta_{\text{eff}}(t) E(\theta)}$ is also displayed in Fig.~\ref{fig:Abs_f}. 
It is important to stress that this agreement is nontrivial, but it can only be \textit{approximate}. If $\abs{f_{\beta}(\theta, t)}$ was exactly thermal, then the $\beta$ dependence in Eq.~\eqref{eq:reproducing_Poisson_kernel} at $t=\frac{\pi}{2 \omega}$ would factorize, since $\beta_{\text{eff}}(\frac{\pi}{2 \omega}) = 0$. Explicit checks show that this is not the case, see for example $f_{\beta}(\theta, t)$ at $\qq=0$, given in Eq.~\eqref{eq:f_beta_for_q=0}.  

A limit where the Gibbs form becomes exact is the Gaussian regime, $\lambda \rightarrow 0$ and $\sqrt{\lambda} \beta \ll 1$. In this case, the $\qq$-oscillator algebra can be approximated by the algebra of an ordinary harmonic oscillator. For a harmonic oscillator, the piece proportional to $\kk$ in the exponent in Eq.~\eqref{eq:t_evolved_TFD} only provides a pure phase $f_{\beta}(E, t) \sim e^{- \frac{\beta}{2} \cos(\omega t) E + i \frac{\beta}{2} \sin(\omega t) E } $ and $\beta_{\text{eff}}(t) = \beta 
\cos(\omega t)$. That we find $\abs{f_{\beta}(\theta, t)}$ to be thermal far from this regime reflects the more general role of $\kk$ in generating phases in the energy eigenbasis. The underlying mechanism is likely the singularity in $K(\theta, \theta^{\prime}, t)$, when $\theta-\theta^{\prime} \sim \omega t$ which is present for all $\qq$. In summary, the analysis so far shows that $\ket{\beta(t)}$ describes, physically, thermal states in the broader sense of thermodynamical ensembles as discussed in Sec~\ref{sec:general_non_ergodic_dynamics}. In addition, it is quite remarkable that by time evolution, the state $\ket{\beta(0)}$ undergoes non-trivial time evolution while remaining instantaneously close to a time-dependent thermal state.   

The return amplitude also has the explicit integral representation 
\be 
A(t) = \frac{(\qq, \qq)^{2}_{\infty}(\tilde{q}(t)^2, \qq)_{\infty}}{ (2 \pi)^2 Z(\beta)}\int^{\pi}_{0} \int^{\pi}_{0}   e^{\frac{ \beta \cos \theta_1}{\sqrt{1-\qq}}} e^{\frac{ \beta \cos \theta_2}{\sqrt{1-\qq}}} \frac{ (e^{\pm 2 i \theta_1}, \qq)_{\infty} (e^{\pm 2 i \theta_2}, \qq)_{\infty}}{ ( \tilde{q}(t) e^{  \pm i \theta_1  \pm  i\theta_2 }, \qq )_{\infty} } d\theta_1 d\theta_2 \label{eq:survival_amplitude_explicit_pochammer}  ,
\ee 
Comparing with Eq. (6.11) of \cite{Berkooz_2018}, we find that Eq.~\eqref{eq:survival_amplitude_explicit_pochammer} is formally similar to a matter two-point function evaluated at $t=0$, with time-dependence instead entering through the complex weight $\tilde{q}(t) = \lim_{z\rightarrow1^{-}} z e^{-i \omega t}$. For comparison, a matter operator composed of fermion strings of size $s$, with dimension $\Delta=\frac{s}{p}$ gets a factor of $\tilde{q} = \qq^{\Delta}$ for each intersection with a chord. The final answer is the sum over all intersections indexed by the chord number. $A(t)$ becomes the generating function for the \textit{spread complexity} distribution \footnote{Spread complexity is the mean position in Krylov space \cite{Parker_2019, Balasubramanian_2022} defined as $\bra{\psi(t)} \hat{n}_0 \ket{\psi(t)}$, where $\hat{n}_0$ is the Krylov number operator with $\hat{n}_0 \ket{\psi(0)}=0$. The definition extends in a straightforward when $\ket{\psi(0)}$ stands for an operator, represented as a state in the Hilbert space of operators. The interpretation and the behavior of spread complexity in various regimes are discussed in \cite{Rabinovici_2021, PhysRevE.111.014218, Erdmenger_2023, Avdoshkin_2024, Chakraborty:2024avd, Caputa_26, Angelinos:2025drf,Review_Nandy_2025} for a non-exhaustive list. } of the time-evolved thermofield double on analytically continuing $\beta$ to $\beta \pm i \tilde{t}$ in \eqref{eq:survival_amplitude_explicit_pochammer}. Here $\tilde{t}$ is an independent parameter labeling the time evolution $e^{-i \frac{1}{4} (\hat{H}_L + \hat{H}_R) \tilde{t}}\ket{\beta}$, and in defining spread complexity, we used $\ket{0}$ as the reference state, treating both the Lorentzian and Euclidean time evolution on even footing. In the JT limit, this amplitude measures the length distribution of the growing Einstein-Rosen bridge \cite{Susskind:2014rva, Lin_2022, Ambrosini:2024sre, Heller_2025, Aguilar-Gutierrez:2025pqp, Aguilar-Gutierrez:2025hty}.   

In terms of the microscopic operator $\hat{S}$, $A(t)$ can be interpreted as the generating function for $\hat{S}$ in the initial state $\ket{\psi}$:
this of interest when $\ket{\psi(0)}$ is of the form $\ket{O} \equiv \hat{O} \ket{0}$, \textit{i.e.} defined by an operator $\hat{O}$ acting on $\ket{0}$ without moving it out of the scar subspace. In this case, the amplitude captures the operator size distribution of $\hat{O}$. The TFD state $\ket{\beta}$ is an example with $\hat{O} = e^{-\frac{\beta}{2} \HH }$ and \cite{Qi_2019} gives a Euclidean path integral computation of $A(t)$ in the two-stage large $p$ SYK model. The key observation is that the operator insertion $e^{-i \mu \hat{S}  t}$ twists the boundary conditions for the Euclidean path integral. The generating function $A(t)$ is determined by the twisted fermion two-point function and the latter, in the $t \rightarrow 0$ limit becomes the thermal Euclidean green's function Eq.~\eqref{eq:large_p_length}. This describes the answer for $A(t)$ in the $\lambda \rightarrow 0$ limit with $\beta$ held fixed. 

\begin{figure}[t!]
    \centering
    \includegraphics[width=0.99\linewidth]{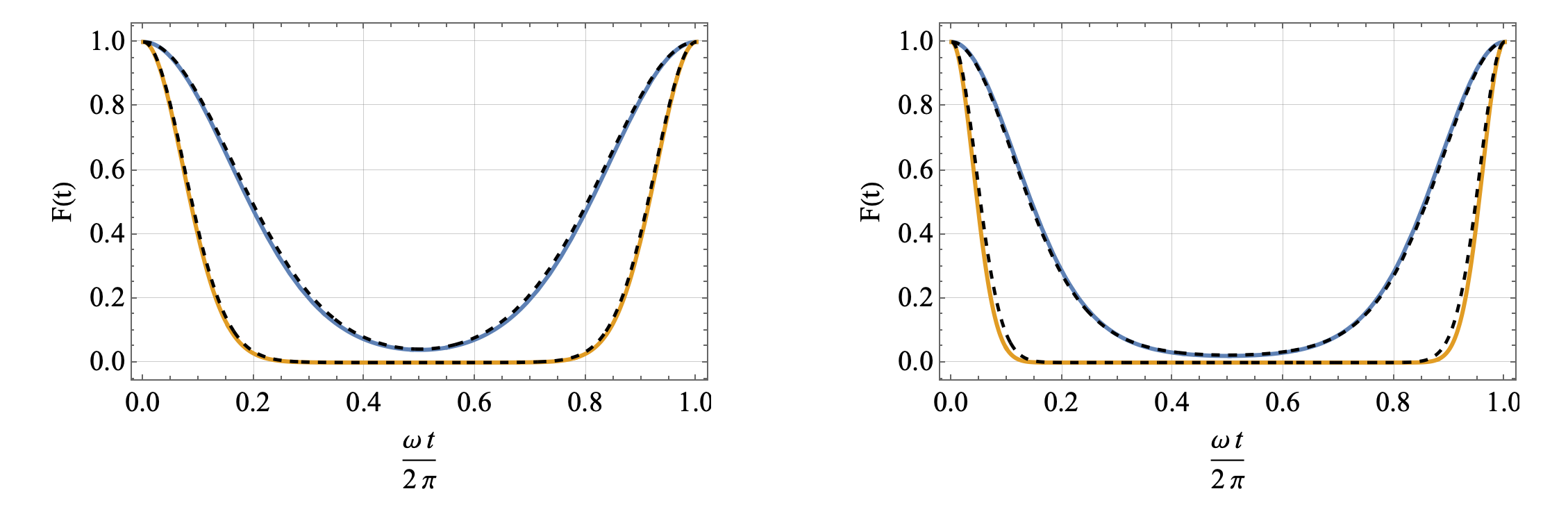}
    \captionsetup{justification=Justified}
    \caption{The return probability $F(t)$ for $\qq=0.1$ (left) and $\qq=0.9$ (right), with $\beta=2$ (blue) and $\beta=10$ (orange). The dashed lines are fits with approximate analytical predictions: for the $\qq=0.1$ plot, we use the predictions coming from the $\qq=0$ case, computed analytically in Eq.~\eqref{eq:survival_amplitde_q=0}; while for the  $\qq=0.9$ plot, we use as an analytical prediction the coherent state approximation computed in Eq.~\eqref{eq:coherent_state_survival_probability}. For this latter case, we used $z = -\frac{1}{2} E(\beta)$ and we see that the agreement with Eq.~\eqref{eq:coherent_state_survival_probability} worsens on increasing $\beta$. }
    \label{fig:survival_probability_inDSSYK}
\end{figure}

Both $A(t)$ and $F(t)$ can be approximated using special limiting values of $\qq$ and $\beta$. As we touched upon before, when $\lambda$ is small and when $\sqrt{\lambda}\beta$ is also small, then $\ket{\beta}$ is similar to a coherent state of an \textit{ordinary} harmonic oscillator, which means that the coefficients $c_n$ are approximately Poisson distributed. A saddle point calculation supporting this is found in \cite{Ambrosini:2024sre}. In practice, we found the validity of this behavior to extend to a range of $O(1)$ values of $\lambda$ and $\sqrt{\lambda} \beta$, in the sense that the low order moments of $\hat{n}$ agree with the coherent state result.  Another tractable limit is that of small $\qq$. Both the return amplitude and the return probability resemble the $\qq=0$ answer, which is discussed in Sec.~\ref{sec:RMT_limit}, in a $\beta$ independent way see Fig.~\ref{fig:survival_probability_inDSSYK}. 
\subsection{Coherent States}
\label{sec:coherent_states}
As already mentioned in Section \ref{sec:validity_of_scars}, another set of initial states that show interesting non-ergodic dynamical features is formed by the set of the coherent states $\ket{z}_{\qq}$ of the $\qq$ oscillators. We are now aiming to provide more details about these states and their dynamical features. 

The states $\ket{z}_{\qq}$ can be thought as a generalization of the $\qq=1$ ordinary coherent states, since they satisfy the defining property $ \aa \ket{z}_{\qq} = z \ket{z}_{\qq} $, with $z$ being a complex number (as we will discuss in a moment, we will need to restrict the maximum value of $\abs{z}$). In the chord basis, $\ket{z}_{\qq}$ takes the explicit expression
\be 
\ket{z}_{\qq}= \frac{1}{\sqrt{\mathcal{N}_{\qq}(\abs{z}^2)}}\sum^{\infty}_{n=0} \frac{z^n}{\sqrt{[n]_{\qq} !}} \ket{\nn},
\ee
with the function $\mathcal{N}_{\qq}$ being the $\qq$ exponential, defined as 
\be \mathcal{N}_{\qq}(v) = \sum_{n} \frac{v^n}{[n]_{\qq}!}  =  \sum_{n} \frac{ (v (1-\qq))^n }{(\qq, \qq)_{n}} = \frac{1}{(v(1-\qq), \qq)_{\infty}}, \  v \in \mathbb C .
\ee
To understand better the profile of $\ket{z}_{\qq}$ in the chord basis, we first note that because $(\qq, \qq)_n$ converges for large $n$, the decay of the weight $\abs{c_n}^2$ is controlled by $|z^2(1-q)|^n$. The moments of the number distribution can be computed by taking derivatives like 
\be ~_{\qq}\bra{z} \hat{n} \ket{z}_{\qq} = v \partial_{v} \log \mathcal{N}_{\qq}(v) |_{v=\abs{z}^2} \ee
and similarly for the higher moments. An alternative, simpler, way of seeing the weights in the chord basis is by taking 
\be ~_{\qq}\bra{z} \aa^{\dagger} \aa \ket{z}_{\qq} =  ~_{\qq}\bra{z} [\hat{n}]_{q} \ket{z}_{\qq} = \abs{z}^2  \ee 
Upon expanding in either $n$ or $q$, we see that $\abs{z}^2 \approx \bra{z} \hat{n} \ket{z}$ for small $z$ for any $q$ (reproducing the property of standard coherent states), and the range of validity of $z$ grows as $\qq \rightarrow 1$, as expected.  

As already mentioned, the value of $\abs{z}$ must be restricted to be sufficiently small. This necessity is two-fold. On the one hand, we aim to consider regimes where the \textit{full} dynamics $e^{-i \hat{H}(\mu) t} \ket{z}_{\qq}$ can be approximated with the simplified expression $ e^{-i \hat{n} \omega t} \ket{z}_{\qq}$. As already mentioned, this requirement is equivalent to the condition $\mu= \omega/p$, with $\omega \in \mathbb R$ and $O(1)$. On the other hand, we notice that  $\ket{z}_{\qq}$ is only normalizable for  $\abs{z}<(1-\qq)^{-\frac{1}{2}}$. For finite $\lambda$, the second condition is more restrictive than the first one. Therefore, we get that the dynamics of the (normalizable) state $\ket{z}_{\qq}$ takes the usual form expected for a coherent state:
\be 
e^{- i \hat{H}(\mu) t} \ket{z}_{\qq} = \ket{z e^{-i \omega t}}_{\qq} \, . \label{eq:coherent_state_dynamics}  
\ee
Given the evolution of Eq.~\eqref{eq:coherent_state_dynamics}, the return amplitude can be computed explicitly to give 
\be  \frac{1}{\mathcal{N}_{\qq}(\abs{z}^2)} \sum_{n} \frac{\abs{z}^{2n} e^{-i n \omega t}}{[n]_{\qq}!}  = \frac{(\abs{z}^2(1-\qq), \qq)_{\infty}}{(\abs{z}^2(1-\qq) e^{i \omega t}, \qq)_{\infty}}, \ee 
from which a closed expression for the return probability can be obtained in the limit $\lambda \rightarrow 0$ to be 
\be 
F(t)=e^{-4 \abs{z}^2 \sin^2(\omega/2) t^2}. \label{eq:coherent_state_survival_probability} 
\ee 
As anticipated in Section \ref{sec:validity_of_scars}, the dynamics of the coherent state $\ket{z}_{\qq}$ (in the limit of a non-normalizable state) can be viewed as the motion of a particle moving rigidly and periodically on the space of the DSSYK eigenstates. To show this feature, we need to consider the expectation value of $\HH$ and its conjugate, as well as their variances.  Let us start by considering the expectation value $\langle \HH \rangle$  in the state $\ket{z e^{-i \omega t}}_{\qq}$. Given its property of behaving as the position operator for the deformed harmonic oscillator, we get readily
\be
\langle \HH \rangle =  \frac{2 \sJ \abs{z} \cos(\omega t + \alpha ) }{\sqrt{\lambda}},
\label{eq:coherent_H}  
\ee 
where $z=\abs{z} e^{-i \alpha}$. Similarly, its conjugate variable is given by 
\be 
\langle \kk(t) \rangle = - \frac{ \sJ \abs{z}}{\sqrt{\lambda} } \sin(\omega t + \alpha). \label{eq:coherent_k} 
\ee 
The second moments are given by 
\be \frac{\lambda}{ \sJ^2} \expval{\HH^2(t)} = \abs{z}^2 (1+\qq) + 2 \abs{z}^2 \cos(2 (\omega t + \alpha)) + 1.  \label{eq:coherent_2nd_moment} \ee 
with a similar expression for $\expval{\kk^2(t)}$. Therefore, the variances are given by 
\be  
\Delta \HH ^2 = 4 \Delta \kk^2 = \frac{ \sJ^2}{\lambda} (1-\abs{z}^2 (1-\qq) ) = \frac{\sJ^2}{\lambda} \expval{\qq^{\hat{n}}},  \label{eq:coherent_variances} 
\ee 
where the last equality confirms that the variances are non-negative (as expected) but that they can go arbitrarily close to $0$. As a consistency check, we see that these states saturate the standard lower bound for uncertainty 
\be 
\frac{1}{2} \abs{\bra{z} [ \HH, [\HH, \hat{n}] ]  \ket{z}} = \frac{\sJ^2}{\lambda} ( 1-\abs{z}^2 (1-\qq) ).  
\label{eq:coherent_uncertainty}
\ee 
Now, let us consider the limit $z \rightarrow (1- \qq)^{-\frac{1}{2}}$ -- so that the variances in Eq.~\eqref{eq:coherent_variances} go to zero -- and set $\omega=\lambda$ and $\alpha=0$ for clarity. In this case, we see that Eq.~\eqref{eq:coherent_H} becomes
\be 
\langle \HH(t) \rangle = \frac{2 \sJ \cos \lambda t }{\sqrt{\lambda (1-\qq) } } ~~~~ \langle \kk(t) \rangle = \frac{-\sJ \sin (\lambda t )}{ \sqrt{\lambda (1-\qq) }}.  \label{eq:coherent_states_as_eigenstates}
\ee 
By comparing Eq.~\eqref{eq:coherent_states_as_eigenstates} with the expression of Eq.~\eqref{eq:spectrum_of_DSSYK} describing the spectrum of $\HH$ we finally arrive at the wanted result: by time evolution, the coherent state describes a configuration where the system starts from the zero-momentum $\kk=0$ DSSYK ground state at $t=0$ and then proceeds to evolve through the manifold of DSSYK eigenstates with increasing energy at a rate parameterized by the momentum $\kk(t)$ (the parameter $\lambda t$ is identified with the spectral angle $\theta$). Let us remark that the limit $z \rightarrow (1- \qq)^{-\frac{1}{2}}$ is meaningful in this context, since $\td$ is sent to infinity despite the chord number diverging. \footnote{This can be seen simply from that the divergence of $\expval{\hat{n}^2}$ is as $(z- (1-\qq)^{-1/2} )^{-2}$, and that it does not depend on $p$ which we send to infinity first. } The non-normalizability of these states is familiar from scattering theory. Notice that, considering $\alpha \neq 0$ just results in an inessential shift in the initial conditions, or equivalently, a shift in the spectral angle from which the particle begins moving. The fact that the variance of $\kk$ also vanishes, is related to a second sense in which $\kk$ plays the role of momentum, in that it labels the energy eigenstates through a dispersion relation.  We will return to this point in Sec.~\ref{sec:continuum_limit}.  

The DSSYK Hamiltonian in the chord basis can be interpreted as a scattering Hamiltonian with the Liouville potential. The eigenvalues are determined by the asymptotic region, where the potential $\qq^{\hat{n}} = e^{-l}$ vanishes. This implies that for all states with $\expval{l} \rightarrow \infty$, $H$ acts like the free Hamiltonian. The nontrivial observation here is that the Liouville potential measures the variances of $H$ and $k$ evaluated on the $\qq$ coherent states, which vanish when we take this limit for $z$. Because these are coherent states, the uncertainty is time-independent and Eq. \eqref{eq:coherent_H}, Eq. \eqref{eq:coherent_k} describe the motion of a perfectly localized particle on a spectrum. We will revisit these states for $\lambda \rightarrow 0$, and their relevance to the Liouville Hamiltonian in Sec \ref{sec:continuum_limit}.  
\subsection{Matrix Elements and Approximate Quantum Error Correction Code in DSSYK}
\label{sec:error_correction}
In Sec.~\ref{sec:general_non_ergodic_dynamics} we discussed the time evolution of observables in the scar subspace as described by  Eq.~\eqref{eq:Heisenberg_EOM}. In the DSSYK context, this carries through in just the same way for observables whose action connects chord states with different chord numbers, with $\HH$ and $\kk$ being examples. Additionally, in Sec.~\ref{sec:general_non_ergodic_dynamics} we also examined the implications for local observables assuming ETH. Due to the double scaling limit however, few-body local observables trivialize in the chord states. A good framework for understanding this is through the lens of quantum error correction. The scar subspace spanned by the chord states forms an \textit{approximate} quantum error correction code. This, along with the fact that the chords states are approximate eigenstates, results in trivial dynamics for local operators in the scar subspace for the time scales that we consider.  

We consider $\bra{\psi(t)} \hat{O} \ket{\psi(t)}$ for local operators and expand
\be 
\bra{\psi(t)} \hat{O} \ket{\psi(t)} = \sum_{n} \abs{c_n}^2 O_{nn} + \sum_{m \neq n} c^{*}_{m} c_n O_{mn}  e^{i (E_m - E_n) t}  \label{eq:spectral_representation_DSSYK_op}  
\ee 
where $O_{mn} = \bra{\textbf{m}} \hat{O} \ket{\nn}$. Notice that the off-diagonal matrix elements $O_{mn}$ in Eq.~\eqref{eq:spectral_representation_DSSYK_op} \textit{vanish} for \textit{any} ``light'' $\hat{O}$ which is a linear combination of a Majorana string $\Psi_{I,s}$ (suppressing the $L,R$ indices since they do not matter) with length $s\lesssim p$. In other words, for $\hat{O} = \sum^{s=r}_{s=1} \sum_{|I|=s} M_{I} \hat{\Psi}^{L}_{I}$.
\be  
\bra{\textbf{m}} \Psi_{I,s}  \ket{\nn} \approx 0~~\text{for}~~ s \leq r ~~\bra{\textbf{m}} \hat{O} \ket{\nn} \approx 0  \label{eq:off_diag_matelems} 
\ee 
where $r$ is a $O(p)$ cutoff. This can be seen by observing $\Psi^{s}_{I}$ acting on any state can only change its size by $\pm s$. The sum has to vanish because the size distribution of $\ket{\nn}$ falls off exponentially away from the mean $\sim n p$. 
However, the chord rules imply a much stronger statement, \textit{i.e.} that even for ``heavy'' operators (which means operators having size scaling with $p$), only \textit{specific} choices of $\hat{O}$, suitably joining the chords in a chord diagram, can be non-vanishing. 

The diagonal matrix elements $\bra{\nn } \hat{O}^1_{L}  \hat{O}^2_{R} \ket{\nn}$ are evaluated using one sided traces like $\tr(H^{k_1} O^{1} H^{k_2}  O^{2}   )$. The diagonal elements of one-sided observables like $\hat{O}=\hat{O}_{L}$ with $\hat{O}^2_R= \mathbb{I}$ vanish. This is because the Hamiltonian chords are paired among themselves, and the inner product simplifies to $\tr(O)$ which vanishes due to the Majorana algebra. Operators like $\hat{O}^1_{L}  \hat{O}^2_{R} $ will be non-vanishing if $\hat{O}^1_{L}$ and $\hat{O}^{2}_{R}$ share identical fermion strings across the two copies. This can be seen using linearity on $i^s \bra{\overline{n} } \Psi^{L}_{I_1, s}  \Psi^{R}_{I_2,s} \ket{\overline{n}}$ with $I_1 = I_2$. We notice that $\Psi_{I_1,s}$  needs to be moved passed $n$ Hamiltonian chords within the trace for the contraction. If the size $s$ of $\Psi_{I_1,s}$ scales with $p$ as $\Delta= \frac{s}{p}$, then we can use Poisson statistics for each intersection, and get an average factor of $e^{-\frac{2 s p n }{N}} = e^{-\Delta \lambda  n} = e^{-\Delta l}$. This is an instance of a matter chord insertion among Hamiltonian chords, presented for a Majorana string, instead of the usually considered unit normalized random operators.   

Even when $s$ scales sub-linearly in $p$, the expectation value can be expanded as a power series in $\frac{s p}{N}$ using Eq.~\eqref{eq:fermion_intersection_statistics} to count intersections. Accounting for the intersections with $n$ chords gives $1 - \frac{n s p}{N}$ to leading order in $1/N$, matching the $\Delta \rightarrow 0$ result of the matter operator answer. The chord number is measured by the bilinear operator $\hat{S}$, but note that the latter is \textit{sum} over all $N$ flavors. 

This brings us to an interesting conceptual perspective on the kinematics of chord states. They represent an approximate quantum error correction code because they asymptotically satisfy the Knill-Laflamme conditions \cite{Knill_Laflamme_1997, B_ny_2010, Brand_o_2019}  for the error operators $\hat{\Psi}_{I}$
\be 
\bra{\textbf{m}} \Psi^{L}_{I_1}  \Psi^{R}_{I_2} \ket{\nn} = \delta_{mn} \delta_{I_1, I_2} \big( 1 - \frac{2 n s p}{N}  + O(N^{-2} ) \big) \label{eq:knill_laflamme} 
\ee 
with corrections controlled by $2nsp/N = \Delta l$. We would have an exact quantum error correction code if the $1/N$ corrections were absent. Extrapolating to finite but $N\gg p \gg 1$ but $\lambda \rightarrow 0$, we would roughly have a $[N,  O(\log \frac{N}{p} ), O(p)]$ code in standard qubit error correction notation. The code parameters are written approximately, because of the statistical nature of the construction. The tradeoff between distance and rate of the code has a suggestive interpretation in terms of chord weights. For example, we could consider the approximate code to be defined by the cutoff $ \Delta l <\epsilon$, fixing $N$ and $p$. Enhancing the code distance at the cost of the rate corresponds to choosing smaller $l$, which allows for larger $\Delta$. Conversely, improving the rate at the expense of $\Delta$ corresponds to taking larger values of $l$. 

This DSSYK result has an interesting physical interpretation in terms of matter insertions in the wormhole. $\Delta$ is interpreted as the conformal dimension of the matter fields, and the diagonal matrix element $e^{-\Delta l}$ is what one obtains for matter insertions on opposite side of the wormhole. Eq.~\eqref{eq:knill_laflamme} implies that there exist reliable recovery channels \cite{Knill_Laflamme_1997, B_ny_2010} for empty wormhole states in the presence of light matter insertions. The only difference is that generally the matter fields are taken to be unit normalized, but random sum of fermion strings of fixed size. The same conclusion holds for such operators due to linearity. This code may not have much practical value due to its nonlocal nature, but it would be interesting to study its inherent stability, due to the statistical nature of the chord construction and to examine its relevance to holography.      

Upon introducing some spread in energy for $\ket{\nn}$, the spectral decomposition  \eqref{eq:spectral_representation_DSSYK_op} in terms of chord states remains valid for $t \sim O( \frac{1}{\mu \sqrt{\lambda}} )$. This time scale diverges when we pick $\mu$ proportional to $1/p$. This implies that the expectation values remain time-independent for long times due to the scar subspace being an approximate error correction code. This is a manifestation of the non-thermal nature of the approximate eigenstates $\ket{\nn}$. The vanishing of the off-diagonal matrix elements in Eq.~\eqref{eq:off_diag_matelems} is a kinematic feature independent of the location of $\ket{\nn}$, controlled by $\mu$. 

\section{The limit of vanishing $\lambda$}
\label{sec:continuum_limit}
From Eq.~\eqref{eq:tdep_def}, and the surrounding discussions in Section \ref{sec:validity_of_scars}, it is evident that we can take $\lambda \rightarrow 0$ and preserve the revivals, as long as we consider states with finite support in the chord basis. The fact that $\Lambda_n$ in Eq.~\eqref{eq:size_variance_and_mean} gets suppressed with $\lambda$ is related to the dominance of commutative effects in local models \cite{Erd_s_2014} which was studied in \cite{Berkooz:2024ofm, Almheiri_2024}. In such cases, the dynamics can be truly mapped to that of an ordinary quantum harmonic oscillator. Examples include the coherent states discussed in Sec \ref{sec:coherent_states} with finite $\abs{z}$. 

Notice that taking $\lambda \rightarrow 0$ leads to extensive in $N$ divergences in Eqs.~\eqref{eq:normalized_chord_hamiltonian} and \eqref{eq:spectrum_of_DSSYK}. This has to be expected on general grounds, since both the energy variance and the maximum energy in this limit tend to behave as in local models, and so they become extensive in $N$,  instead of getting regulated by inverse powers of $p^2$. This is also reflected in the extensive scaling of thermodynamical quantities such as the thermal energy at finite $\beta$. Since $\bra{\beta} \HH \ket{\beta}$ diverges with $N$ we can already deduce that $\ket{\beta}$ must have support on states with divergent chord number $n$. This observation can be put on a more quantitative footing following \cite{Qi_2019,Lin_2023}, relating $\bra{\beta} \hat{S} \ket{\beta}$ to the thermal correlator in a single copy of SYK. As a result, the mean chord number $\langle \hat{n} \rangle$ diverges. We define the renormalized chord number operator as $\hat{l} \equiv \lambda \hat{n}$, and we will call it the \textit{length} operator following \cite{Lin_2022,Lin_2023}, where it is shown that this operator describes the length mode in JT gravity \cite{Harlow_2019}. Given this definition, we obtain that the quantity $l = \bra{\beta} \hat{l} \ket{\beta}$ remains finite as $\lambda \rightarrow 0$:   
\be 
l = \lambda \bra{\beta} \hat{n} \ket{\beta} = - 2 \log \cos \frac{\pi v}{2},  \label{eq:large_p_length} 
\ee 
where we introduced the $v$ parametrization of the inverse temperature $\beta$ defined in \cite{Maldacena_16} as
\begin{equation}
    \frac{\pi v}{\beta} \equiv \cos \frac{\pi v}{2}.
    \label{eq:v_parameter_definition}
\end{equation}
In Eq.~\eqref{eq:large_p_length}, the RHS can be obtained by matching on to the answer from two-stage large $p$ limit, see also \cite{Goel:2023svz}. The fluctuations of $\hat{l}$ are known to be small \cite{Qi_2019, Lin_2023}, so we can treat $\langle  \hat{n}^2\rangle  \approx \langle  \hat{n} \rangle^2 = \frac{l^2}{\lambda^2}$ and from \eqref{eq:tdep_def} we arrive at 
\be 
\td/t_{\text{rev}} \approx \frac{\sqrt{\lambda} p}{l},  \label{eq:bounds_finite_N}
\ee 
showing that a \textit{sufficient} condition for the revivals to persist is to consider thermofield doubles with short enough length $l$, although we should keep in mind that there may be further damped oscillations for time scales longer than $\td$.  We expand Eq.~\eqref{eq:large_p_length} for small $\beta$, getting $l \sim \beta^2$ and find that there are a finite number of cycles for $\beta \lesssim \lambda^{\frac{1}{4}} \sqrt{p}$. When $p$ is held fixed as a function of $N$, this vanishes. To find controlled revivals in that limit, we could define $\beta = \lambda^{\alpha} \tilde{\beta}$ with $\beta \rightarrow 0$,  $\alpha \geq \frac{1}{4}$ and $\tilde{\beta}$ held fixed. The case with $\alpha=\frac{1}{2}$ has been studied in \cite{Almheiri_2024}, and it happens to precisely be the scaling for which $\bra{\beta} \HH \ket{\beta}$, entering in Eq.~\eqref{eq:Heisenberg_H}, becomes $O(1)$ instead of extensive. In this case $\ket{\beta}$ coincides with the notion of the $\lambda \rightarrow 0$ limit of the coherent state studied in \ref{sec:coherent_states} as we alluded to before in \ref{sec:revival_dynamics_DSSYK}. 
Notice that we cannot infer the breakdown of the approximate correspondence between $\hat{n}$ and $\hat S$, Eq.~\eqref{eq:approx_chord_number_op} if the demand for small values of the renormalized chord number $l$ is not met.
As a matter of fact, revivals are sensitive to energy differences, and Eq.~\eqref{eq:bounds_finite_N} reflects that the chord states have a spread in size comparable to the size difference between consecutive chord states. Both the difference and the spread in size are made small by picking $\mu= \frac{\lambda}{p}$\footnote{This discussion remains unchanged if we had additional $O(1)$ factors in the definition of $\mu$. The precise choice is for the sake of simpler notation.} so that $\td= \frac{p}{\sqrt{\lambda} l}$. This is a continuum limit where the first cycle of oscillation gets stretched out over a divergent $t_{\text{rev}} = \frac{2 \pi}{\lambda}$ time scale. In this situation, the return amplitude $A(t)$ goes from being a Fourier series with coefficients $c_n$, to being a Fourier transform of the coarse-grained length distribution. 

We will now show that in this limit, \textit{any} wavepacket moves ballistically on the spectrum and due to the dispersion relation, its energy grows quadratically. Let us first notice that the length $\hat{l}$ takes continuous values and generates the time evolution. It follows that time evolution \textit{coincides with momentum evolution}. This momentum refers to the operator that is conjugate to the length $\hat{l}$, defined as 
\be \hat{a}^{\dagger} =  e^{i \lambda \tilde{k}},  ~~~ \tilde{k} \equiv  i \partial_{l}. \ee 
With these conventions $[ \hat{l}, \tilde{k}  ] = -i$. In the Schwarzian limit, $\tilde{k}$ is the momentum labeling energy eigenstates, and also the wormhole momentum in JT gravity \cite{Harlow_2019}. The momentum evolution can be better understood by writing the Heisenberg equations (setting $\sJ=-1$) for operators in the subspace
\be \tilde{k}(t) = \tilde{k} + t, \ee 
\be  \HH(t) = -\frac{ \big( e^{-i \lambda ( \tilde{k} + t) } \sqrt{1-e^{-\hat{l}}} +  \sqrt{1-e^{-\hat{l}}} e^{i \lambda ( \tilde{k} + t) } \big)}{\sqrt{\lambda (1-\qq )}}~~~,~~~ \label{eq:continuum_time_dependent_H}\ee 
\be \kk(t) = \frac{i}{2} \frac{ \big( e^{-i \lambda ( \tilde{k} + t) } \sqrt{1-e^{-\hat{l}}} -  \sqrt{1-e^{-\hat{l}}} e^{i \lambda ( \tilde{k} + t) } \big)}{\sqrt{\lambda (1-\qq )}}. \label{eq:continuum_time_dependent_k}  \ee 
Therefore time evolution leads to the linear growth of the momentum $\tilde{k}$ for any state in the subspace. $\tilde{k}$ labels the spectral angle for eigenstates similar to Eq.~\eqref{eq:coherent_states_as_eigenstates}, so that the energy eigenvalues in term of $\tilde{k}$ read
\be 
E(\tilde{k} ) = -\frac{2 \cos \lambda \tilde{k}  }{\sqrt{\lambda (1-\qq) }}.  \label{eq:E_as_function_of_tildek} 
\ee 
Recalling the discussion in Sec. \ref{sec:coherent_states} around Eq.~\eqref{eq:coherent_states_as_eigenstates}, we interpret the motion of the localized particle in this regime in terms of the linearly growing momentum $\tilde{k}$. We have not directly used that $\lambda$ is infinitesimal so far except to define $\tilde{k}$. Now we take the continuum limit by taking $\lambda \rightarrow 0$, and to make contact with the Schwarzian regime, we take $l \rightarrow \infty$ with $l+2 \log(\lambda)$ held fixed and find a time-dependent Liouville Hamiltonian on-shell: 
\be 
\HH(t) = \lambda \Big( \tilde{k}(t)^2 + \frac{e^{-\hat{l}}}{\lambda^2} \Big)  -\frac{2}{\lambda} \label{eq:time_dependent_Liouville} \ee 
\be  \kk(t)  = \tilde{k}(t) = \tilde{k}(0) + t  \label{eq:two_momentum_relation} 
\ee 
Eq. \eqref{eq:two_momentum_relation} shows that in this limit, the momentum $\tilde{k}(t)$ and $\kk(t)$, measuring the ``speed'' of the particle moving on the spectrum are equivalent. Another consequence of the continuum limit, in contrast to the revival dynamics presented in Sec \ref{sec:revival_dynamics_DSSYK} is that the particle only moves in one direction, instead of bouncing back and forth. That is \eqref{eq:time_dependent_Liouville}, implies that the dynamics explores one-sided states with \textit{only increasing energies}. A similar picture holds when we keep $\hat{l}$ finite, which amounts to dressing the RHS of Eqs.~\eqref{eq:time_dependent_Liouville} and \eqref{eq:two_momentum_relation} with some time-independent functions of $\hat{l}$. 

\subsection{Comments on the holographic interpretation}
\label{sec:holographic_interpretation}
There are different holographic dictionaries for DSSYK \cite{Lin_2022,Susskind:2022bia, Blommaert_2024, Verlinde2025}. In all cases, the holographic dual describes the dimensional reduction of a gravitational theory, where the reduced degrees of freedom are described by the two-dimensional phase space of the $\qq$ oscillator in DSSYK. The $\lambda \rightarrow 0$ limit becomes the semiclassical limit, as evident from the way $\lambda$ appears in the expression for $\rho(\theta)$ or $Z(\beta)$. There is mounting evidence that DSSYK describes anti-de Sitter space at low-energies, and de Sitter space at low $\beta$.   

An important point is the role of $\HH$ as the generator of time evolution in the physical system. The maximal scrambling of $\HH$ is crucial for identifying SYK as a system capturing the dynamics of black holes \cite{Maldacena_16,Maldacena_2017}. Instead, here, we are talking about a sector of two copies of SYK where the dynamics is non-chaotic by design. 

This is best understood in terms of the holographic duality between SYK and AdS$_2$ JT gravity coupled to matter. The two copies of SYK are naturally identified with two Lorentzian boundaries of near-AdS$_2$ space. The lack of chaos for this model is conceptually similar to the case of Maldacena-Qi wormhole \cite{maldacena2018eternaltraversablewormhole}. There, interactions between matter fields and the Schwarzian modes make the system dual to global AdS$_2$ at low temperatures. Instead, here we consider a non-equilibrium setting, with the initial conditions specified by $\ket{\psi(0)}$, such that all the dynamics is being sourced by the interaction term in Eq.~\eqref{eq:H_tot_def}. The effective time evolution operator is very similar to the instantaneous coupling studied in \cite{Maldacena_2017} for the wormhole teleportation protocol in SYK. We will address the gravitational dynamics of the near-AdS$_2$ boundaries in Lorentzian signature in future work \cite{Chakraborty_2026}.  

Instead, we can try to directly address the states explored by the non-ergodic dynamics in terms of the bulk path integral preparing them, in the spirit of \cite{Kourkoulou:2017,Goel_2019}. Since it is a lot easier to constrain $\abs{f_{\psi}(\theta, t)}$ from the revival dynamics, it is easier to approach this question from the perspective of the Euclidean action. 

For example, $\ket{\beta}$ corresponds to the classical solution of the Euclidean path integral corresponding to a hyperbolic disc of circumference proportional to $\beta$. From the discussion in Sec ~\ref{sec:revival_dynamics_DSSYK}, we know that $\ket{\beta(t)}$ is not a TFD. Nonetheless, we found evidence in Sec~\ref{sec:revival_dynamics_DSSYK} that $\abs{ f_{\beta}(\theta, t) }$ remains thermal with an effective $\beta_{\text{eff}}(t)$ for a range of $\lambda$ and $\beta$; we will conjecture that this continues to hold true in the vanishing $\lambda$ regime while zooming in on the low-energy spectrum. This $\beta_{\text{eff}}(t)$ can be computed by comparing Eq.~\eqref{eq:time_dependent_Liouville} with the thermal expectation values
\be 
\langle H \rangle_{\beta_{\text{eff}}(t)} =  \frac{1}{ \lambda \sJ} \frac{(2 \pi)^2}{\beta^2_{\text{eff}}(t)} = \frac{1}{\lambda \sJ} \frac{(2 \pi)^2}{\beta^2} + \lambda \sJ^3 t^2 \implies \beta_{\text{eff}}(t) =\frac{2 \pi  \beta}{\sqrt{4 \pi ^2 + \sJ^4 \beta^2  \lambda^2
   t^2 }}. \label{eq:effective_beta_lambda_0}
\ee
where $\beta$ specifies the TFD at $t=0$ and factors of $\sJ$ were reintroduced to clarify the dimensional analysis. The operational meaning of $\beta_{\text{eff}}(t)$ is in line with the discussion presented in Sec ~\ref{sec:general_non_ergodic_dynamics}. This $\beta_{\text{eff}}(t)$ will determine the outcome of instantaneous local measurements on one copy. The fact that $\beta_{\text{eff}}(t)$ becomes nearly independent of $\beta$ at large $t$, is a consequence of the linear growth in momentum $\tilde{k}(t)$ for small $\lambda$. Because the unperturbed classical solution is a hyperbolic disc, it is reasonable to associate with $\ket{\beta(t)}$, a disc geometry with a time-dependent defect. The circumference of the disc, in units of boundary value of dilaton, is given by $\beta_{\text{eff}}(t)$--which implies that the disc is shrinking.

It is also interesting to consider the revival dynamics of Sec ~\ref{sec:revival_dynamics_DSSYK} in the light of its relation to de Sitter as proposed in \cite{Narovlansky:2023lfz, Verlinde2025}. Though it is not clear how to justify the embedding of Eq.\eqref{eq:H_tot_def} in the de Sitter context, the subspace in which we find revival dynamics is precisely the one studied in \cite{Narovlansky:2023lfz} to relate two-point functions in de Sitter and DSSYK. \cite{Verlinde2025} relates the spectral angle in $\ket{\theta}$ to the conical deficit angle caused of 3d Schwarschzild de-Sitter spacetimes. Since the revival dynamics described in Sec ~\ref{sec:revival_dynamics_DSSYK} moves the system along the spectrum, it has a clear operational meaning in terms of a process scanning through different semiclassical geometries.  
\section{Finite-size revivals}
\label{sec:finite_size_revivals}
The results obtained in the previous Sections have characterized the highly non-ergodic dynamics, induced by the Hamiltonian in Eq.~\eqref{eq:total_hamiltonian_mu-shifted}, experienced by the states living in $\Hkryl$, in the double-scaling limit.
While the double-scaling limit is certainly interesting in view of the holographic interpretation of the model and for providing an analytically tractable regime, it is also interesting to investigate the robustness of the non-ergodic dynamics in the more standard situation of a finite $N$, $p = 4$, two-sided SYK model. In this section, we will perform this analysis. 

\begin{figure}
    \centering
    \includegraphics[width=0.47\linewidth]{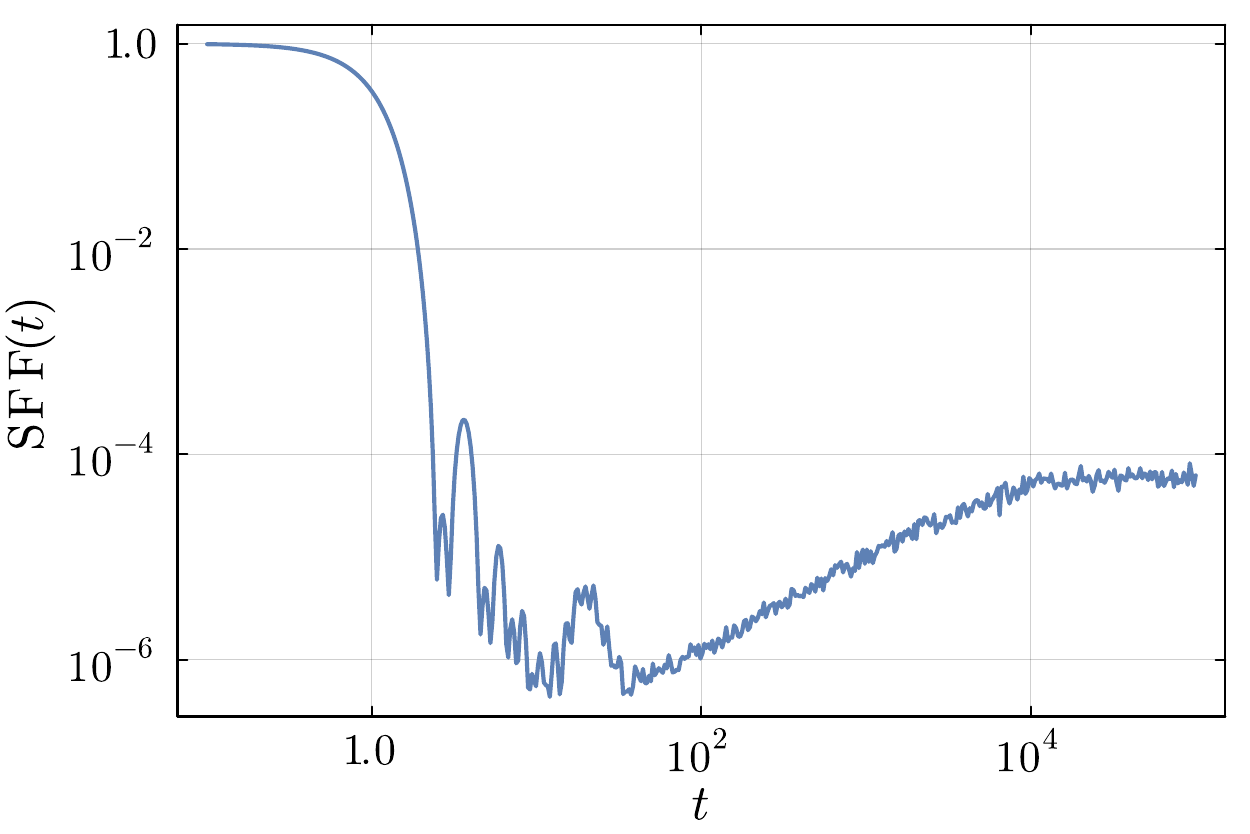}
    \captionsetup{justification=Justified}
    \caption{The behavior of the SFF, as a function of $t$, for the case $N = 28$ and $\mu = 0.2$. The ``dip-ramp-plateau'' structure, indicative of the chaotic character of the Hamiltonian Eq.~\eqref{eq:H_tot_def}, is evident.}
    \label{fig:SFF}
\end{figure}

As a preliminary check, we need to confirm numerically that the model, as a whole, is ergodic. To this end, we have computed the so-called spectral form factor (SFF), which is a standard probe for chaotic behavior and agreement with the RMT predictions \cite{guhr1998random-matrix, Cotler_2017}. Let us remind that the SFF is, by definition, the modulus squared of the analytically continued partition function. In formulas, given the Hamiltonian of Eq.~\eqref{eq:H_tot_def}, we define the following quantity
\begin{equation}
    \label{eq:SFF_def}
    \mathrm{SFF}(t) \equiv \frac{1}{D^2} \sum_{n,m}e^{i (E_m - E_n) t},
\end{equation}
where $D$ denotes the dimension of the full Hilbert space, while $E_m$ are the energy levels of the full Hamiltonian in Eq.~\eqref{eq:H_tot_def}. The results are then averaged over several realizations of the disorder. The behavior of the averaged SFF, as a function of $t$, is reported in Figure \ref{fig:SFF}. As can be clearly seen from the plot, the averaged SFF shows a robust ``dip-ramp-plateau'' behavior, the latter being a hallmark of RMT, thus confirming that the model defined by Eq.~\eqref{eq:H_tot_def} is generically chaotic.
Having clarified the chaotic nature of the full Hamiltonian, we now move to the investigation of the non-ergodic behavior for the subspace $\mathcal{H}_\mathrm{Krylov}$.
On general grounds, it is clear that the discrepancy between the operator $\hat{S}$ and the chord operator $\hat{n}$, as expressed by Eq.~\eqref{eq:size_variance_and_mean}, is more and more important when considering states $\ket{\nn}$ with increasing values of $n$, with the extreme case of the infinite temperature TFDs, $\ket{0}$, which is an eigenstate of $\hat{S}$ for any (finite) value of $N$ and $p$. Given these considerations, it is natural to consider the time evolution of TFD states, $\ket{\beta}$, with $\beta$ not too large. These states will likely have large support on states $\ket{\nn}$ with $n$ small, so that the approximate non-ergodic dynamics will more likely be preserved. 
\begin{figure}
    \centering
    \includegraphics[width=0.47\linewidth]{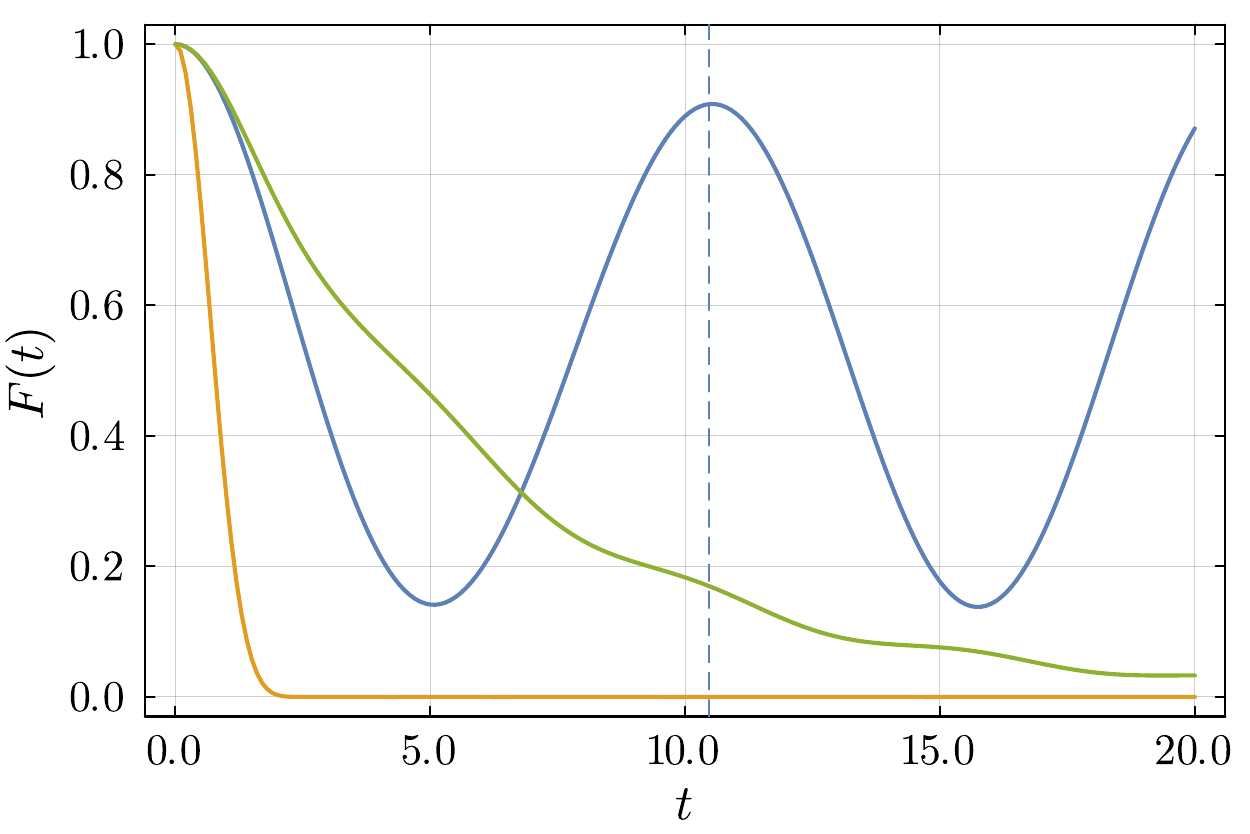}
    \includegraphics[width=0.47\linewidth]{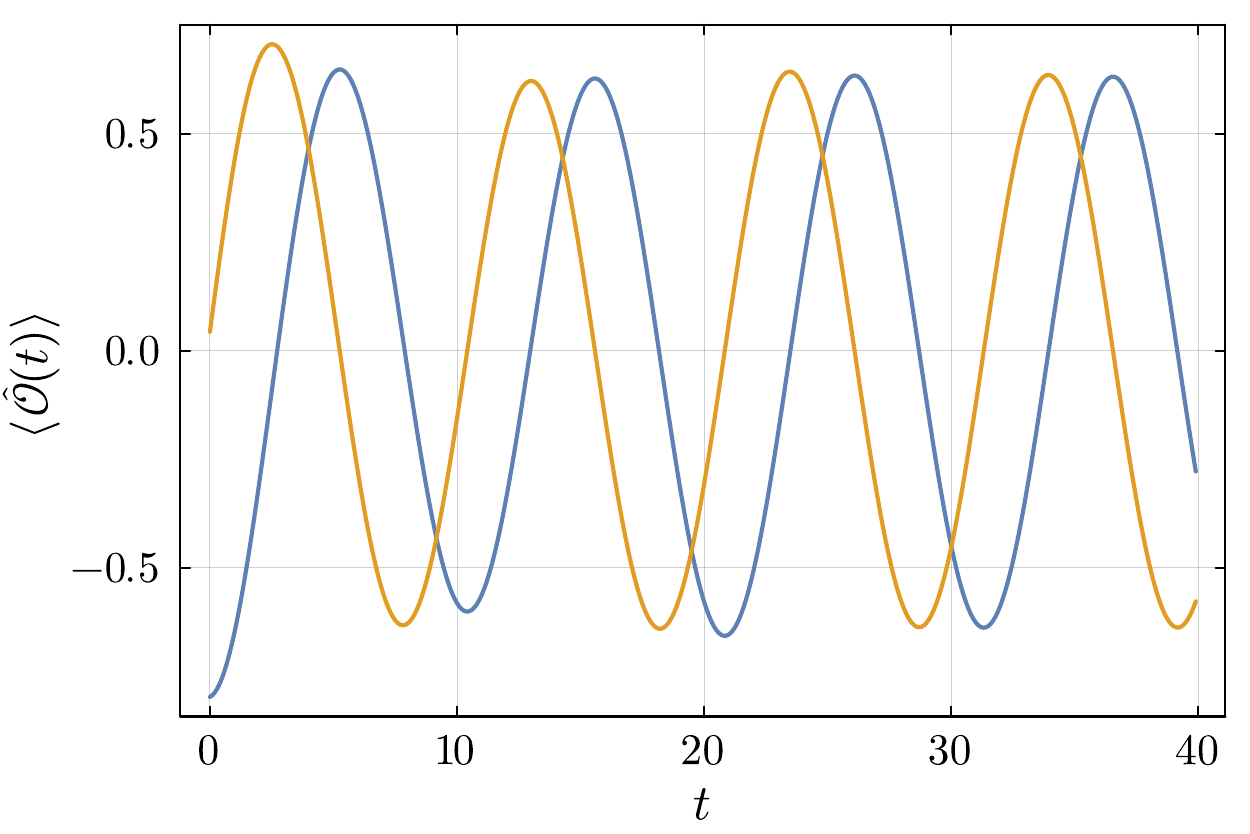}
    \captionsetup{justification=Justified}
    \caption{Left panel: the absence/presence of non-ergodic dynamics for $N = 16$, $\mu = 0.15$. The state $\ket{\beta}$, with $\beta = 1.0$, shows robust non-ergodic dynamics, as clearly visible from the revivals of the return probability $F(t)$ (blue line). On the contrary, both a random initial state living in the full Hilbert space (orange line), as well as a random initial state living in $\Hkryl$ (green line), show a rapid decay of $A(t)$, although the decay of the state living in $\Hkryl$ is much slower in time. The vertical dashed line is the value of $t_\mathrm{rev}$ as predicted by Eq.~\eqref{eq:t_rev}. Right panel: the time evolution, $\langle \hat{\mathcal{O}}(t) \rangle$, with $\hat{\mathcal{O}}$ being $\hat{H}_L$ (blue curve) and $2 \hat{k}_0$ (orange line), for the state $\ket{\beta}$, with $\beta = 1.0$ and $\mu = 0.15$.}
    \label{fig:revival_dynamics_N32_beta_1_mu_0d15}
\end{figure}

The expectations above are numerically tested, for the specific case of $N = 16$ and $\mu = 0.15$, in Figure \ref{fig:revival_dynamics_N32_beta_1_mu_0d15}. As we can see from the temporal evolution of the return probability $F(t)$, the TFD state $\ket{\beta}$, with $\beta = 1.0$, shows a robust revival dynamics, with revivals bringing $F(t)$ up to values of order $0.9$. Another important consistency check of the validity of the double-scaling predictions is that the period of the revivals is in excellent agreement with the expected $t_\mathrm{rev}$ computed using Eq.~\eqref{eq:t_rev}. On the other hand, we observe a very different behavior when we consider the dynamics of a state initialised at random in the full Hilbert space, for which the return probability rapidly goes to zero. This decay is, of course, in agreement with the generically ergodic character of the model. Finally, we see that by taking a random state living in $\Hkryl$, the revivals are absent, but the decay is much slower compared to the decay of a generic random state. This result is in agreement with the expectation that, at finite $N$ and $p$, only a small portion of the states $\ket{\nn}$ can be considered as approximate eigenstates of $\hat{S}$, thus making the non-ergodic dynamics of a generic state in $\Hkryl$ not robust. Another confirmation of the validity of the predictions coming from the double scaling limit can be obtained by studying the time evolution of the expectation values of $\hat{H}_0(t)$ and $2 \hat{k}_0(t)$ in the state $\ket{\beta}$, with $\beta = 1$ and $\mu = 0.15$. The results are reported in the right panel of Fig.~\ref{fig:revival_dynamics_N32_beta_1_mu_0d15}, which shows an excellent agreement with the prediction of Eq.~\eqref{eq:Heisenberg_H}  and Eq.~\eqref{eq:Heisenberg_k}, with the phase shift between the two curves which reproduces the theoretical expectations.
\begin{figure}
    \centering
    \includegraphics[width=0.47\linewidth]{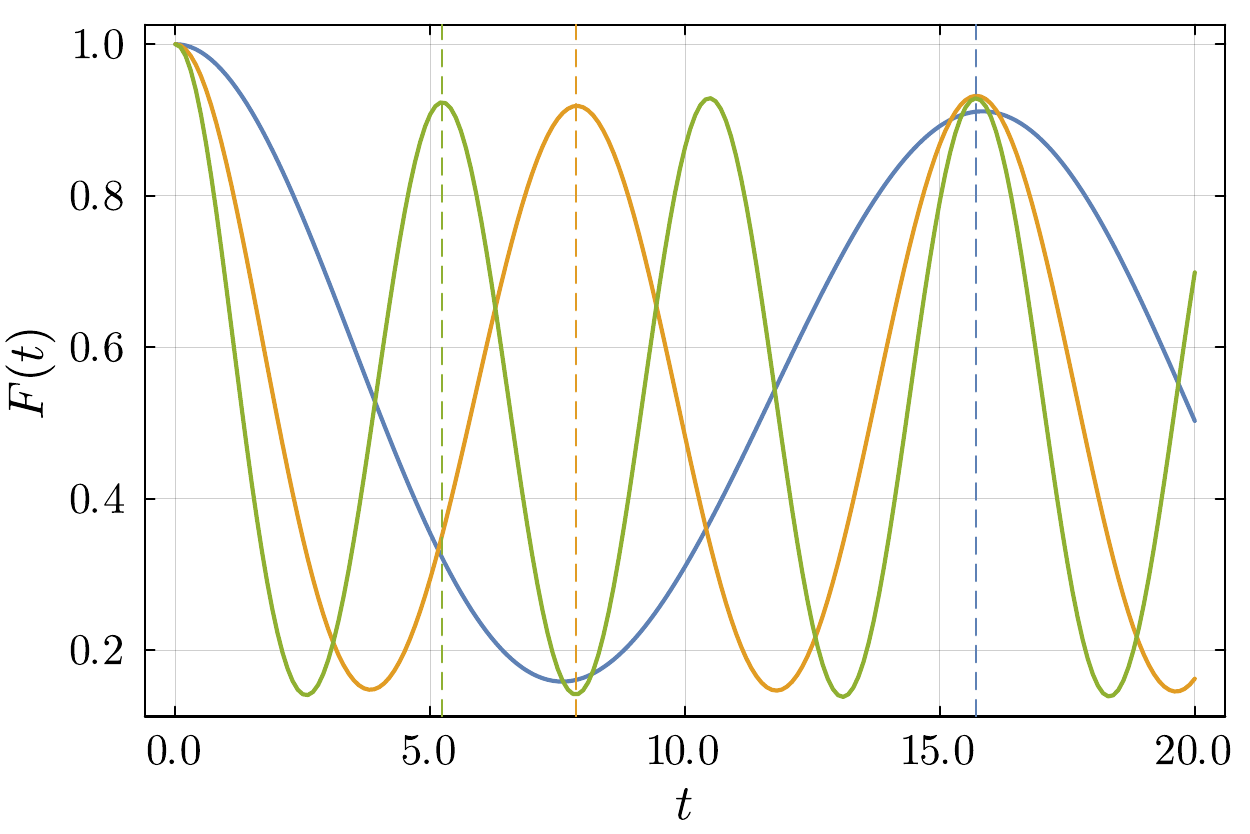}
    \includegraphics[width=0.47\linewidth]{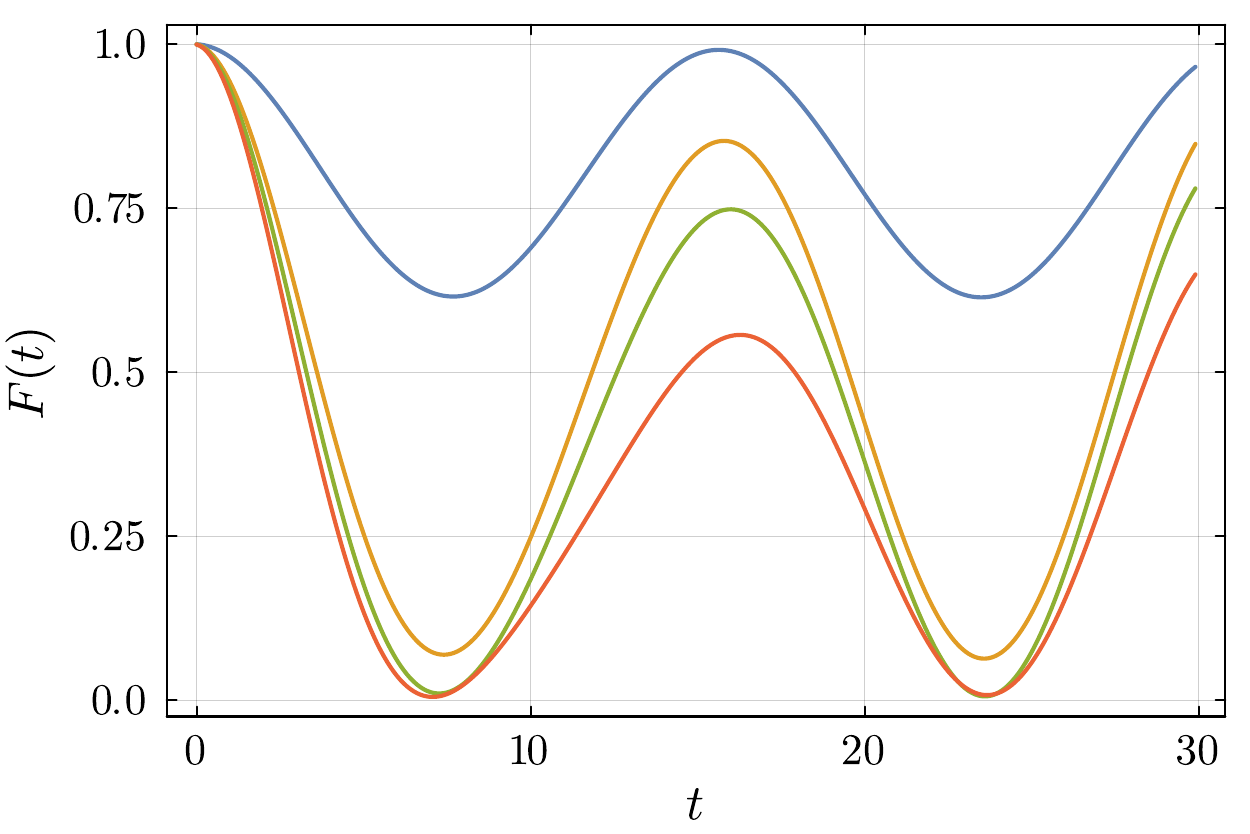}            \captionsetup{justification=Justified}
    \caption{Left panel: the revival dynamics for values of $\mu = 0.1$ (blue line), $0.2$ (orange line), and $0.3$ (green line), at $N =16$ and $\beta = 1$. The vertical lines show the agreement with the analytical predictions of Eq.~\eqref{eq:t_rev}. Right panel: the return probability $F(t)$, computed for $N = 16$ and $\mu=0.1$ at varying values of $\beta$: $0.5$ (blue line), $1.2$ (orange line), $1.5$ (green line), and $2.0$ (red line).}
    \label{fig:robustness_revival_dynamics}
\end{figure}
The features of the non-ergodic dynamics just described are robust to changes in the value of the coupling $\mu$. This stability is manifestly visible in the left panel of Figure \ref{fig:robustness_revival_dynamics}, where the revivals are clearly visible for $\mu = 0.1, \ 0.2, \ 0.3$, and where the agreement with the predictions of Eq.~\eqref{eq:t_rev} is excellent. On the other hand, as expected, the revival dynamics is highly sensitive to the value of $\beta$, as shown in the right panel of Fig.~\ref{fig:robustness_revival_dynamics}. For $\beta ~0.5$ we see that the revival probability $F(t)$ oscillates mildly, reaching the minimum value of $F \approx 0.6$. This is consistent with the idea that for values of $\beta$ small, the state $\ket{\beta}$ will have a large overlap with the infinite temperature TFD $\ket{0}$, which is an eigenstate, thus making the revival dynamics mild. On the other hand, we see that for $\beta \approx 2.0$, $F(t)$ reaches a minimum value which is approximately zero but, at the same time, it poorly reconstructs the initial state, reaching a maximum $F(t) \approx 0.5$. So, the sweet spot is represented by values of $\beta \approx 1.0$, for which $F(t)$ reaches minimum values close to $0$ and then reaches a maximum of $\sim 0.75$ or larger.

All in all, the results of this Section show the robustness of the approximate non-ergodic dynamics for finite values of $N$ and $p$. Remarkably, the predictions of the double-scaling limit are largely reproduced at finite $N$ and $p$, as shown by the excellent agreement of the period of the revivals computed numerically with the analytical prediction of $t_\mathrm{rev}$ as in Eq.~\eqref{eq:t_rev}.
\section{Discussion}
\subsection{Summary of Results}
We presented a new construction of non-ergodic scar-like dynamics using two perfectly anticorrelated copies of a system, which are then coupled in a specific way. The perfect anticorrelation between the two uncoupled copies creates an exponentially large number of zero energy eigenstates when the two systems are decoupled. These zero-energy eigenstates form a subspace of the full Hilbert space spanned by the diagonal eigenstates in the full system.
We then define an interaction term which preserves this subspace, and further introduces a grading structure in it, thus making the eigenstates belonging to different sectors having equally spaced energies. In Sec.~\ref{sec:general_construction}, we fleshed out the construction for the specific case in which the subspace is the Krylov space  $\mathcal{H}_\mathrm{Krylov} = \text{span}\left\{H^j_{L} \ket{0} = H^{j}_{R} \ket{0} = \HH^{j} \ket{0}\right\}$ where the seed, $\ket{0}$ is the infinite temperature thermofield double of the decoupled system. The interaction term, leading to the grading structure is the Krylov number operator, $\hat{n}_0$. The scarred subspace contains physically relevant states, including the purifications of thermal states of the decoupled system, which includes the thermofield double. More generally, the states belonging to this subspace can be viewed as \textit{equilibrium states} for one of the two copies in the absence of interaction.

Although this is a general construction, which holds irrespective of the details of the Hamiltonians governing the two decoupled systems, there is a large degree of universality in the resulting dynamics of the subspace which is elaborated in Sec.~\ref{sec:general_non_ergodic_dynamics}. This universality is a simple consequence of the fact that the decoupled Hamiltonian $\HH$ acts like the position operator for a harmonic oscillator, where the generator of time evolution in the subspace is the number operator $\hat{n}_0$. The dynamics of $\ket{\psi(t)}$ in the scar subspace is summarized as a wavepacket moving along the spectrum of the decoupled left/right system. The reduced density matrix for $\ket{\psi(t)}$ restricted to one copy is an equilibrium state for the decoupled system and the entanglement entropy $S(t)$ between the two copies precisely determines the spread of the time-dependent wavepacket. If the wavepacket is narrow enough, and assuming eigenstate thermalization, the time-dependent expectation values in the scar subspace can be obtained from \textit{thermal} values for the decoupled system. Each copy then sees an effective time-dependent temperature, determined by the instantaneous value of the one-sided energy. 

Though this formalism is powerful and general, the explicit form of number operator $\hat{n}_0$ can be quite messy when expressed in terms of the microscopic operators. Therefore, a crucial step to support the  physical relevance of the construction is to find concrete examples of interactions where the number operator $\hat{n}_0$ could be embedded.  The connections between $\hat{n}_0$ and operator size, and the existence of simple operators which measure the ``size'' of operators acting on $\ket{0}$ \cite{Qi_2019, Schuster_2022, Parker_2019} suggest  the existence of at least approximate interaction terms in which $\hat{n}_0$ could be embedded. As a concrete model, we take two copies of double-scaled SYK, which can be analytically solved using chord diagrams. For two-sided DSSYK, the chord number operator  coincides with the Krylov number operator $\hat{n}_0$. The chord number operator also has an approximate microscopic realization as a bilinear term pairing a fermion in the left copy with one in the right copy \cite{Lin_2022}. The total Hamiltonian including the interaction term is related to the Maldacena-Qi Hamiltonian by changing the relative sign. For this system, $\HH$ is explicitly known in the chord basis in terms of raising and lowering operators satisfying a $\qq$ deformed oscillator algebra.  

We study this dynamics in DSSYK in Sec.~\ref{sec:revival_dynamics_DSSYK} in a regime where errors due to approximation can be neglected. For DSSYK, we found that the time evolution of a TFD $\ket{\beta}$, retains approximately thermal subsystems, with the time-dependence that can be practically encoded into an effective $\beta_{\text{eff}}(t)$. Decoupled energy eigenstates emerge as a limit of $\qq$ coherent states and realize the motion of a sharply localized particle on the spectrum. The discussion for local observables is quite different in DSSYK, since their dynamics trivializes in this limit, a property that can be understood in the language of approximate quantum error correction code, as we discussed in Sec.~\ref{sec:error_correction}. 

The interaction operator $\hat{n}_0$ continues to have approximate eigenstates in $\hat{H}_\mathrm{Krylov}$, when moving out from the strict double-scaling limit of SYK, \textit{i.e.} when $\lambda \rightarrow 0$. In the $\lambda \rightarrow 0$ limit, the non-ergodic dynamics for TFDs and for the eigenstates of the uncoupled systems ceases to describe revivals, though remaining highly non-ergodic. Due to the inherent approximations, we get a continuum limit where the wavepacket accomplishes only the first cycle of oscillation over a divergent timescale. Restricted to low energies, this corresponds to the dynamics of a time-dependent Liouville Hamiltonian, as we comment on in Sec.~\ref{sec:continuum_limit}.
The chord states are not well-defined for finite $N$ and $p$. Nevertheless, the chord diagrams still give the leading order (in $1/N$) answer for the moments, and there continue to be states in $\mathcal{H}_\mathrm{Krylov}$ which are approximate eigenstates of the interaction operator. The asymptotic revival dynamics, including the energy oscillations, continue to persist for states with small enough size. We presented numerical evidence for the revivals and the remnants of the non-ergodic dynamics, for finite values of $N$ and $p$, in Sec.~\ref{sec:finite_size_revivals}. The results show a remarkable degree of agreement with the double-scaling predictions, therefore reinforcing the confidence that this mechanism provides a robust way to embed non-ergodic dynamics into an otherwise thermalizing mesoscopic system. 
\subsection{Outlook}
This work leaves many avenues for further research and investigation, with a broad range of extensions touching different topics. 
\begin{enumerate}
    \item \textit{Explicit realizations} \\ 
Probably the most immediate question to address is to find other systems where the construction described in Section \ref{sec:general_construction} can be realized, either approximately or exactly, perhaps with a system that is easier to realize experimentally. A related question is to find the general conditions under which a ``reasonable'' realization of $\hat{n}_0$ can be found in terms of a microscopic operator.

From an experimental point of view, it is quite difficult to engineer two perfectly anticorrelated copies of an all-to-all interacting disordered model like SYK, although see \cite{Lantagne_Hurtubise_2020, Haenel_2021} for proposals in this direction. On the bright side, there already exist experimental demonstrations of rainbow-scar like construction even in the presence of disorder \cite{Dong_2023}. On a platform like the one used in \cite{Dong_2023}, it is a matter of tuning $\HH$, and finding an interaction term acting like $\hat{n}_0$ on the equilibrium subspace. 

The mirror transformation discussed in Section \ref{sec:general_construction} provides flexibility in the choice of signs of coupling and the related discussion about the flow of time. For example, in $p=6$ SYK the two copies have Hamiltonians with the same sign of coupling in Eq.~\eqref{eq:H_tot_def}. Alternatively, one can consider models enjoying a particle-hole symmetry, \textit{i.e.} systems for which all energies appear in pairs with opposite signs, for which the sign choice becomes immaterial. 

Another interesting direction to explore is to understand the connection between the framework developed in this paper and the notion of \textit{Lindblad many-body scars}, introduced in \cite{garcia-garcia2025lindblad} to describe the presence of many-body scars in dissipative setups. A related construction, presented in \cite{Lychkovskiy_2024} maps the identity operator in a dissipative system to a thermofield double, which acts as a scar in the doubled system with a modified Hamiltonian. Clearly, these constructions show substantial similarities and just as many differences, and it will be important to explore in detail the connection between them.

\item \textit{Probing Size Distributions} \\ 
For the SYK example that we presented, we have seen that the probability amplitudes capture the distribution of size and complexity. In more general settings, we assume that $\hat{n}_0$ operator can be embedded in a microscopic operator measuring the operator size. It would be interesting to devise protocols to measure complexity distribution of equilibrium states (i.e of $e^{-\beta \hat{H}_L}$ for example) using this dynamics. 

It would also be interesting to better understand the connections between our protocol and the closely related gravitational and size-based teleportation protocols discussed in \cite{Maldacena_2017, Schuster_2022}. For these applications, understanding the effect on ``matter'' insertions on the thermofield double state is important. For sufficiently small insertions to the TFD, some of the scar dynamics could be preserved for times much smaller than the scrambling time, but we leave a quantitative analysis of this to future work.  

\item  \textit{Equilibrium States of the decoupled system} \\ 
A striking feature of the dynamics that we studied is that, after turning on the coupling, equilibrium states of the decoupled system develop an evolution which is \textit{entirely} through other equilibrium states of increasing (or decreasing) energies. At the same time, the system efficiently scrambles generic states, \textit{i.e.} states that, from the point of view of a single copy, \textit{are not} equilibrium states. As we have discussed, under additional mild assumptions, this generates an ensemble of thermal states with time-dependent temperatures. This gives a lot of predictive power for measurement outcomes, since there exist powerful techniques for thermal calculations. 

Another, potentially highly compelling, feature is that, from the point of view of a single copy, the dynamics is also \textit{quasi-static} in the thermodynamics sense, although it is realized in \textit{finite-time}, and could be leveraged as a tool to study the thermodynamics of the decoupled system in a controlled way, especially in experimental applications. 
Two obvious possible applications of this peculiar dynamics are in quantum thermodynamics \cite{vinjanampathy01102016, Binder:2018rix, campbell2025roadmap}. The first application relies on the property that the evolution of the full system is happening in finite time, but the states visited from the point of view of a single copy are equilibrium states. Therefore, this type of evolution holds huge potential to realize a \textit{shortcut to adiabaticity} protocol \cite{shortcuts}. It will be extremely interesting to explore in the future the applications of this idea. Also, we have seen that by time evolution thermal states (so, \textit{passive states}) evolve into states that are characterized by population inversions. This kind of evolution can be thought of as a charging protocol of a quantum battery \cite{alicki2013entanglement,campaioli2024colloquium}, and it will be interesting to see this possibility at work in concrete protocols.

\item \textit{Full Sector} \\ 
We discussed the non-equilibrium dynamics of the Hamiltonian defined by Eq.~\eqref{eq:H_tot_def} for states with no matter insertions, \textit{i.e.} states in $\Hkryl$. There are many other interesting aspects of the full system specified by Eq.~\eqref{eq:H_tot_def}. For example, the two-point function in the thermal state of the combined system can be studied and it undergoes a zero temperature phase transition as a function of $\mu$ \cite{Chakraborty_2026}, such that, for $\mu>\mu_c$, the infinite temperature TFD $\ket{0}$ becomes the ground state of the coupled system. Meanwhile, at high temperatures, the Green's functions match those of two copies of decoupled SYK model \cite{Chakraborty_2026}. It would be interesting to understand the sector of the full system Eq.~\eqref{eq:H_tot_def} describing NAdS$_2$ dynamics, and possibly also in the context of de Sitter physics. 

\item  \textit{Connections with Holographic Scars} \\ 
\cite{Milekhin_2024, Dodelson_2022} have studied many-body scars for holographic systems in higher dimensions. Like in our case, these scars are only \textit{approximate} eigenstates. The scars in \cite{Dodelson_2022} describe orbit states in black-hole backgrounds, which means they have volume-law entanglement. 

\item \textit{Applications of Chord algebra} \\ 
The emergent two-sided DSSYK chord algebra with arbitrary matter insertions has a subalgebra that commutes with the chord number operator. This becomes the $SL(2,R)$ algebra of $AdS_2$ isometries in the JT limit. The explicit scar construction relies on grading the subspace annihilated by the boost generator by $\hat{n}$. It would be interesting to construct non-ergodic dynamics using specific sectors that include matter. For example, one could consider other boost-invariant configurations including matter, and still use $\hat{n}$ to split the sectors, such that the total Hamiltonian is still chaotic, analogous to Eq.~\eqref{eq:H_tot_def}. Finally, exploring the error correction code structure of the full chord algebra might advance understanding of the emergence of bulk locality in JT gravity with matter. 
\end{enumerate}
\section*{Acknowledgements}
We thank S.~Aguilar-Gutierrez, M.~Berkooz, G.~Benenti, T.~Iadecola, H.~Lin, J.~Magan , J.~Murugan, and Z.~Papic for interesting discussions on this and related topics. 
DC is supported by the S\~ao Paulo Research Foundation (FAPESP) through the grant 2024/13100-8.
DR acknowledges FAPESP for the ICTP-SAIFR grant 2021/14335-0 and the Young Investigator grant 2023/11832-9.
DR also acknowledges the Simons Foundation for the Targeted Grant to ICTP-SAIFR.

\bibliography{scars}
\appendix
\section{Chord States as approximate eigenstates}
\label{sec:Chords_as_approx_eigenstates}
Here we elaborate on what it means for the chord states $\ket{\nn}$ to be approximate eigenstates of Eq.~\eqref{eq:H_tot_def}. We will work assuming the double-scaling limit and discuss afterwards how these results might be extrapolated to $p$ scaling differently. The chord states are annihilated by $\hat{H}_L - \hat{H}_R$, so their energy distribution is determined by the size distribution of the chords. Only the pairwise intersections of the chord index sets matter, since higher intersections are disfavored in this limit. Each intersection between the chords reduces the size (i.e, $\hat{S}$ eigenvalue) by $2$. The probability of two index sets having $m$ intersections is Poisson distributed as $(\lambda/2)^{m} e^{-\lambda/2}/m!$ for \textit{finite} $m$.  The precise tail distribution can be inferred from \eqref{eq:fermion_intersection_statistics}, which could also be used to directly work in the fixed $p$ regime.

Neglecting details of the exponentially suppressed tails, we can assume a Poisson distribution for each pairwise intersection, and use the fact that the sum of Poisson random variables also has a Poisson distribution. Thus the total probability of intersections is Poisson distributed with mean $\Lambda_n = \binom{n}{2} \lambda$ and we write down an approximation for the weights appearing in   $\ket{\nn}  = \sum_s c^{n}_s \ket{s}$ as 
\be 
\abs{c^{n}_s}^2 =\frac{\Gamma( 1+\frac{n p}{2} )}{\Gamma(1+ \frac{n p}{2}, \Lambda_n/2) }  \frac{\big( \Lambda_{n}/2 \big)^m e^{-\Lambda_n/2}}{m!}, ~~~~ 2m = np - s ~~~    
\label{eq:Poisson_size}
\ee 
where $m \in \mathbb Z^{+}$. Thus the mean energy and the energy variance are 
\be 
\epsilon_n =  \bra{\nn} \hat{H}(\mu) \ket{\nn} \approx \mu( n p -  \Lambda_n ). \label{eq:energy_mean} \ee 
\be \bra{\nn}  \hat{H}(\mu)^2 \ket{\nn} - \bra{\nn} \hat{H}(\mu) \ket{\nn}^2 = \mu^2  (\bra{\nn} S^2 \ket{\nn} - \bra{\nn} S \ket{\nn}^2 ) \approx  \mu^2 \Lambda_n \label{eq:energy_variance}. 
\ee
The energy variance suffices to bound the overlap between $\ket{\nn}$ and the true energy eigenstates, which we call $C_n(E)$. For example, we could use the conservative, but general Chebyshev bound
\be 
\int_{\abs{E-\epsilon_n > r \mu^2 \Lambda_n}} \abs{C_{n}(E)}^2 dE  \leq \frac{1}{r^2} 
\ee 
to estimate the weight of the tails of $\ket{\nn}$ in the energy eigenbasis. Because of the infinite system limit, we take $C_n(E)$ to be a continuous function but absorb any factors of the density of states.  

Using \eqref{eq:Poisson_size} and \eqref{eq:energy_variance}, we can conclude for the return amplitude of an arbitrary $\ket{\psi(0)} = \sum_n c_n \ket{\nn}$
\be 
\fid(t) \approx \sum_{n} \abs{c_{n}}^2  \bra{\nn} e^{-i \hat{H}(\mu) t} 
\ket{\nn}\ee 
or in other words, because of the small spread of $\ket{\nn}$ in the true eigenbasis of $\hat{H}(\mu)$, even with time evolution, the chord states of different chord number will remain orthogonal.
The range of validity of
\be 
\fid(t) \approx \sum_{n} \abs{c_{n}}^2 e^{-i \mu n p t}  \label{eq:amplitude_DS_limit} 
\ee 
can be inferred on general grounds from the properties of the Fourier transform. Each exponential in \eqref{eq:amplitude_DS_limit} dephases at a time scale that is given by the inverse width  \eqref{eq:energy_variance} about the mean \eqref{eq:energy_mean}. For a simple but concrete toy model, we could take each $\ket{\nn}$  to be spread like a normalized rectangular box of length $d_n =  \mu \sqrt{3 \Lambda_n}$ about the mean $\epsilon_n$ for consistency with \eqref{eq:energy_variance} and \eqref{eq:energy_mean}. Thus 
\be 
\fid(t) \approx \sum_n \abs{c_{n}}^2 e^{-i \mu (n p-\Lambda_n) t} \frac{\sin d_n t   }{d_n t } \label{eq:fidelity_model} 
\ee 
Eq. ~\eqref{eq:fidelity_model} has the merit of being able to capture oscillations for $A(t)$ that signifies nonergodic dynamics, and their decay, away from the regime where Eq. ~\eqref{eq:amplitude_DS_limit} holds as good approximation. Using \eqref{eq:fidelity_model}, we can estimate the ``dephasing'' time $t_{\text{dep}}$ by Taylor expanding in $d_n$ and looking for the scales at which Eq. ~\eqref{eq:fidelity_model} can be approximated by \eqref{eq:amplitude_DS_limit}. We can explicitly check by expanding Eq.~\eqref{eq:fidelity_model},  that $t_{\text{dep}} \sim \frac{1}{\mu \sqrt{\sum_{n} \abs{c_n}^2 \Lambda_n  }}$.  Or in other words:
\be 
t_{\text{dep}} \sim \frac{1}{\mu \sqrt{\lambda \expval{n^2}}}, ~~~t_{\text{revival}} \sim \frac{2 \pi}{\mu p},  \label{eq:time_scales} 
\ee 
where $\expval{n^2} \equiv \sum_{n} \abs{c_n}^2 n^2 $. This predicts near perfect revivals for $t_{\text{dep}}/t_{\text{revival}} \sim \frac{p}{\sqrt{\lambda}} =\sqrt{N}$ cycles if we assume that $\expval{n^2}$ converges, and doesn't scale with $p$.  

Setting $c_{n} = \delta_{mn}$, we find $A(t) \approx e^{-i \epsilon_m t} \frac{\sin( d_m t  ) }{d_m t } $.  This tells us that for time-scales $t  \lesssim \frac{1}{\mu  \sqrt{\Lambda_m}}$, the chord state retains phase coherence $e^{-i \HH  t}\ket{\textbf{m}} \approx e^{-i \epsilon_m t} \ket{\textbf{m}}$ to a good approximation. A specific model of $C_n(E)$ has been used in writing \eqref{eq:fidelity_model}, which demonstrates how \eqref{eq:time_scales} comes about. The short time expansion of $\bra{\nn} e^{-i \HH t} \ket{\nn}$ for arbitrary $C_n(E)$ shows the generality of \eqref{eq:time_scales}. The precise form of $\ket{\nn}$ determines the proportionality constants in the definition of $t_{\text{dep}}$. 

Since we assumed the double scaling limit to write \eqref{eq:time_scales}, we could neglect the $\Lambda_n$ term on the right hand side of Eq \eqref{eq:energy_mean} while estimating $\trev$. The deviations in revivals due to $\epsilon_n$ not being evenly spaced will be subleading, relative to the loss of coherence captured by $t_{\text{dep}}$. When extrapolating to the finite $N$ and $p$ case \eqref{eq:energy_mean}, \eqref{eq:energy_variance} continue giving the leading order answer following the discussion in Sec \ref{sec:finite_size_revivals}. 

\section{Chord Basis Details}
\label{sec:Chord_Basis_Details}
We review here the technical details of the chord construction in a self-contained way. The two-sided Hamiltonian has the following representation in the unnormalized chord basis $\{ \ket{n} \}$:
\be \HH = \frac{\sJ}{\sqrt{\lambda}}(a^{\dagger} +  a [\hat{n}]_\qq), ~~~~[\hat{n}]_\qq \equiv \frac{1-\qq^{\hat{n}}}{1-\qq} \label{eq:tridiag} \ee 
in terms of the oscillators defined in \eqref{eq:transfer_matrix_moments} and $[n]_\qq = \frac{1-\qq^{n}}{1-\qq}$ is the $\qq$ integer, such that $[\hat{n}]_{\qq} \ket{n} = [n]_{\qq} \ket{n}$. The inner product compatible with the chord rules presented there is given by 
\be  \langle n | m \rangle = \delta_{nm} [n]_{\qq}! ~, ~~ [n]_{\qq}! \equiv \frac{(\qq, \qq)_{n}}{(1-\qq)^n}  \ee
where $(a, \qq)_{n} = \prod^{i=n-1}_{i=0}(1- a \qq^i)$ is the $\qq$-Pochammer symbol. To describe the action of $H$ on the normalized states $\{ \ket{\nn} \}$, it is useful to introduce the oscillators
\be  \aa  \equiv a \sqrt{[\hat{n}]_\qq} ~~~ \aa^{\dagger} = \sqrt{[\hat{n}]_\qq}a^{\dagger}  \ee 
explicitly, these oscillators have the action
\be \aa \ket{\nn} = \sqrt{[n]_\qq} \ket{\nn-1 } ~~, ~~ \aa^{\dagger} \ket{\nn} = \sqrt{[n+1]_\qq} \ket{\nn +1} \ee 
and their algebra is given by
\be [\hat{n}, \aa] = -\aa, ~~~ [\hat{n}, \aa^{\dagger}] = \aa^{\dagger}, ~~~  [\aa, \aa^{\dagger}]_{\qq}=1, \label{q_def_alg}\ee
where $[A, B]_{\qq} = AB-\qq BA$ is the $\qq$ commutator. The Hamiltonian has the form of a ``position operator'' in the chord number basis
\be 
\HH =\frac{\sJ}{\sqrt{\lambda}} (\aa+\aa^{\dagger}) \label{eq:normalized_chord_hamiltonian_app} 
\ee 
Eq~\eqref{eq:normalized_chord_hamiltonian} is a tridiagonal matrix in the chord basis and can be diagonalized explicitly to find 
\be 
\HH \ket{\theta} = E(\theta) \ket{\theta}, ~ E(\theta)= \frac{2 \sJ \cos(\theta) }{ \sqrt{\lambda(1-\qq)} } \label{eq:spectrum_of_DSSYK_app} ~~  
\ee 
where the energy eigenstates are parameterized by the angle $\theta \in [0, \pi]$. Since the Hilbert space is infinite in this limit, the spectrum is continuous and $E(\theta)$ takes values in  $[ \frac{- 2\sJ  }{ \sqrt{\lambda(1-\qq)} },   \frac{2\sJ  }{ \sqrt{\lambda(1-\qq)} }]$. $\ket{\theta}$ is a state on the two-sided Hilbert space of the form $\ket{E(\theta)}_L\ket{E(\theta)}_R$ where $\ket{E(\theta)}$ is a SYK eigenstate for a single copy. The spectrum is symmetric under $\sJ \rightarrow -\sJ$. This is a consequence of the disorder average. We will send $\sJ \rightarrow - \abs{\sJ}$ in Eq~\eqref{eq:normalized_chord_hamiltonian}. With this choice of sign, the ground state is at $\theta=0$. We will send $\sJ=-\sqrt{\lambda}$ in Eq~\eqref{eq:normalized_chord_hamiltonian_app} to simplify some expressions and match the conventions of \cite{Berkooz_2018}. We should keep in mind that $\beta$ enters the equations in the dimensionless combination $\beta \sJ/\sqrt{\lambda}$.  
The density of states $\rho(\theta)$, normalized so that $\int^{\pi}_{0} \rho(\theta) d\theta=1$ is 
\be  
\rho(\theta) = \frac{1}{2 \pi} (\qq, \qq)_{\infty} (e^{2 i \theta}, \qq)_{\infty} (e^{-2 i \theta}, \qq)_{\infty} = \frac{\sin \theta }{\pi \qq^{\frac{1}{8}}} \vartheta_{1}(\theta, \qq^{\frac{1}{2}} ), \label{eq:rho(theta)} 
\ee
where we are using the Mathematica conventions for the Jacobi $\vartheta_{1}$ function. 
Eq.~\eqref{eq:normalized_chord_hamiltonian_app} together with Eq~\eqref{eq:spectrum_of_DSSYK_app} implies the recursion for $\phi_n(\theta) = \bra{\theta}\ket{\nn} $
\be E(\theta) \phi_n(\theta) = \sqrt{[n]_{\qq}} \phi_{n-1}(\theta)   +   \sqrt{[n+1]_{\qq}} \phi_{n+1}(\theta).   \ee 
which is solved by
\be 
\sqrt{ (\qq, \qq)_n  } \phi_n(\theta) = H_{n}(\cos(\theta)| \qq ) \equiv \sum^{n}_{k=0} \frac{ (\qq, \qq)_n ~ e^{i (n-2k) \theta}}{  (\qq, \qq)_{k}  (\qq, \qq)_{n-k} }  \label{eq:qHermite}
\ee 
the $H_n$ are known as the continuous $\qq$-Hermite polynomial. The completeness and orthogonality relations for the continuous $\ket{\theta}$ basis and the discrete chord basis are summarized as 
\be 
\int^{\pi}_{0} d\theta \rho(\theta) \phi^{*}_m(\theta) \phi_n(\theta) = \bra{\mathbf{m}}\ket{\nn} = \delta_{mn}. 
\ee 
\be 
\langle \theta | \theta^{'} \rangle   = \sum_{n} \phi_{n}(\theta) \phi^{*}_{n}(\theta') = \frac{\delta(\theta- \theta^{'})}{\rho(\theta)}.  
\ee 
The partition function $Z(\beta)$ thus is given by 
\be  Z(\beta) = \bra{0} e^{-\beta H} \ket{ 0 } = \int^{\pi}_{0} d\theta \rho(\theta) e^{-\beta E(\theta) }  \label{eq:partition_function_integral}  \ee 
and the amplitude for finding $\ket{\beta}$ in the $n$th chord state is given by the related matrix element computed in \cite{Berkooz_2018}, \cite{Okuyama_2023} 
\be
\bra{\nn}  \ket{\beta}= \frac{1}{\sqrt{Z(\beta)}}\sum^{\infty}_{r=0} e^{i \pi r}  \qq^{\frac{r}{2} (r +1)} \frac{(\qq, \qq)_{r+n}}{\sqrt{(\qq, \qq)_n} (\qq, \qq)_r} 
     (2r+n+1) \frac{2 \sqrt{1-\qq} }{\beta} I_{2r+n+1}(\frac{ \beta}{ \sqrt{1-\qq}}), \label{eq:tfd_in_chord_basis}
\ee
where $I$ is the modified Bessel function of the first kind. This gives an explicit infinite series representation for $Z(\beta)$, the return amplitude $A(t)$ and related quantities. For fixed $\beta$ and $\qq$, the RHS of \eqref{eq:tfd_in_chord_basis} decays as a function of both the summation index $r$, and the chord index $n$. 
The evolution of $\ket{\psi(t)}$ in the $\ket{\theta}$ basis is tracked using the matrix $\bra{\theta} \ket{\theta^{\prime}(t)} $ which is the Poisson Kernel 
\begin{align}
K(\theta, \theta^{\prime}, t ) &= \sum^{\infty}_{n=0} \phi_n(\theta) \phi_{n}(\theta^{\prime} ) e^{-i n \omega t }  = \sum^{\infty}_{n=0} H_n(\cos \theta|\qq) H_n(\cos \theta^{\prime} |\qq)  \frac{( e^{-i \omega t} )^n}{(\qq, \qq)_{n}}\\ 
&= 
 \lim_{z \rightarrow 1^{-}} \frac{(z^2 
e^{-i 2 \omega t}, \qq)_{\infty}}{(z  e^{i (\theta+\thetap -  \omega t)},z  e^{i (\theta-\thetap -  \omega t)}, z e^{-i (\theta+\thetap + \omega t)}, z  e^{-i (\theta-\thetap + \omega t)}, \qq)_{\infty}}  = \lim_{z \rightarrow 1^{-}} \frac{ (z^2 e^{-2i\omega t}, \qq)_{\infty}}{ ( z e^{ - i \omega t  \pm i \theta_1  \pm  i\theta_2 }, \qq )_{\infty} },  \label{eq:generating_function_series}
\end{align}
where in the first line we used that $\phi_n$ are real. Strictly speaking, the kernel Eq.\eqref{eq:generating_function_series} is only convergent for all arguments when $\abs{z}<1$. On the unit circle, it has isolated singularities due to the denominator vanishing at $\pm \theta  \pm \thetap = \omega t $  but these are integrated over, such that $f_{\psi, \qq}(\theta)$ is well-defined. In practice, for the numerics, we do this by taking the $z \rightarrow 1^{-}$ of the integrals. 
\section{ Random Matrix Theory Limit }
\label{sec:RMT_limit}
The case $\qq=0$ constitutes the random matrix theory limit. We express the semicircle law in terms of $E(\theta)$ and find
\be \rho(\theta) = \frac{2}{\pi} \sin^2(\theta), ~~~  Z(\beta) = \frac{I_{1}(2 \beta )}{\beta}~~~\phi_n(\theta) = \frac{\sin(n+1) \theta}{\sin \theta}, \ee  
which gives thermal energy and thermal entropy:
\be E(\beta) = - \frac{2 I_{2}(2 \beta)}{I_{1}(2 \beta) } ~~~S(\beta)=  - \frac{2 \beta I_{2}(2 \beta)}{I_{1}(2 \beta) } + \log( \frac{I_{1}(2 \beta )}{\beta} ).  
\ee 
Only the first term in the sum over $r$ in \eqref{eq:tfd_in_chord_basis} contributes and
\be 
\bra{\nn} \ket{\beta} =  \frac{2 (n+1)}{\sqrt{Z(\beta)}} \frac{I_{n+1}( \beta )}{\beta}.  \label{eq:c_n_for_q=0}   
\ee 
From this we get the series representation for $f_{\beta}(\theta,t)$ 
\be 
f_{\beta}(\theta,t) = \frac{2}{\beta \sqrt{Z(\beta)}} \sum^{\infty}_{n=0}  (n+1) I_{n+1}( \beta )  \frac{\sin(n+1) \theta}{\sin \theta} e^{- i n \omega t}, 
 \ee
which can be rewritten as
\be f_{\beta}(\theta,t) = \frac{e^{i \omega t}}{i \sqrt{Z(\beta)} \sin \theta } \big[ \mathcal{T}(\theta-\omega t) -\mathcal{T}(-\theta-\omega t)  \big]~~, ~~~ \mathcal{T}(x) =  \sum^{\infty}_{n \geq 1} n  I_{n}(\beta) e^{i n x}.  \label{eq:f_beta_for_q=0}
\ee 
using the expansion,
\be  e^{\beta \cos(x) } = I_0(\beta) + 2 \sum^{\infty}_{n=1}  I_n(\beta) \cos n x,  \ee 
and the three-term recurrence for $I_n(\beta)$
\be  I_{n-1}(\beta) - I_{n+1}(\beta) = \frac{2 n }{\beta} I_n(\beta), \ee 
we find that $\mathcal{T}(x)$ can be written slightly more explicitly as 
\be 
\mathcal{T}(x) =  \frac{1}{2}\Big[ i  \sin(x)  e^{\beta \cos
   (x)}+   I_0(\beta) \cos x  +I_1(\beta)  +  \sin(x) \Im \sum_{n \in \mathbb Z} I_{n}(\beta) e^{i n x} \Big] ,
\ee 
which is combined to get the form 
\begin{equation}
\begin{split}
    f_{\beta}(\theta,t) & = \frac{e^{i \omega t}}{i \sqrt{Z(\beta)} \sin(\theta) } \Big[ I_0(\beta) \sin(\omega t) \sin(\theta) +  \frac{i}{2} ( \sin (\theta - \omega t  ) e^{\beta \cos (\theta - \omega t  ) 
   }+\sin (\theta + \omega t ) e^{\beta \cos (\theta +\omega t  )  } ) \\  
    &- \sum_{n \geq 1}  I_n(\beta) \big( \sin ( n-1) \theta  \sin ( n-1) \omega t -\sin( n+1) \theta \sin ( n+1) \omega t      \big)   \Big]. \label{eq:simplified_f_q0}
\end{split} 
\end{equation}
From both Eqs. \eqref{eq:simplified_f_q0} and \eqref{eq:c_n_for_q=0}, we can find representations of $A(t)$, though they all involve a piece without a standard closed form. The return amplitude is written as  
\be 
A(t) =  e^{i \omega t} ( C(t) + i S(t) ),  \label{eq:survival_amplitde_q=0}
\ee 
where
\be 
C(t) = \frac{ \cos \frac{ \omega t  }{2}  I_1(2 \beta \cos \frac{\omega t }{2})  -\beta  \sin
   ^2 \frac{\omega t}{2}   \big( I_0(2 \beta 
   \cos  \frac{\omega t}{2} ) +I_2(2
   \beta  \cos \frac{\omega t}{2} )  \big) } {I_1( 2 \beta )},   ~~~ S(t) = \frac{4}{I_1(2 \beta)} \sum^{\infty}_{n=0} n^2 I^2_n(\beta) \sin n \omega t. 
\ee 
It is possible to approximate $S(t)$ in a number of different ways, including by approximating the sum by an Euler-Maclaurin integral. The Taylor coefficients for the expansion of $S(t)$ about $t=0$ can be obtained as finite linear combinations of Bessel functions on using the recurrence relation.  Another manageable expansion is the small $\beta$ expansion of $I_n(\beta)$ for which the sum over $n$ can be explicitly performed at each power of $\beta$. The leading term is found to be 
\be S(t) = \frac{i \beta \left(  e^{-i \omega t  } I_0\left(e^{-i \omega t
   } \beta \right) - e^{ i \omega t  }
   I_0\left(e^{i \omega t  } \beta \right)\right)}{2 I_1(2 \beta )} + O(\beta^2).
\ee 
\end{document}